\documentclass[%
 reprint, superscriptaddress,
amsmath,amssymb,longbibliography,aps,prx,
article]{revtex4-2}
\pdfoutput=1
\usepackage{graphicx,multirow}%
\usepackage{amsmath}
\usepackage{gensymb}
\usepackage{bbm}
\usepackage[inkscapelatex=false]{svg}
\usepackage{dcolumn}
\usepackage{bm}
\usepackage{xcolor,colortbl}
\usepackage{color}
\definecolor{blue}{rgb}{  0,  0,    1}
\definecolor{d4blue}{rgb}{  0,  0.4470,    0.7410}
\definecolor{d12orange}{rgb}{0.8500    0.3250    0.0980}
\definecolor{g8yellow}{rgb}{0.    0.6    0.298}
\definecolor{ppurple}{rgb}{0.4940    0.1840    0.5560}

\def\red{\color{red}}

\newcolumntype{k}{>{\columncolor{blue!20}}c}
\newcolumntype{r}{>{\columncolor{blue!10}}c}
\newcolumntype{d}{>{\columncolor{red!10}}c}
\usepackage{comment}

\newcommand{\down}{\downarrow}
\newcommand{\up}{\uparrow}

\usepackage{hyperref}

\begin{document}

\preprint{APS/123-QED}

\title{Gap structure and phase diagram of twisted bilayer cuprates from a microscopic
perspective}

\author{Siddhant Panda}
\affiliation{Department of Physics, University of Florida, Gainesville, Florida 32611, USA}
\author{Andreas Kreisel}
\affiliation{Niels Bohr Institute, University of Copenhagen, DK-2100 Copenhagen, Denmark}
\affiliation{Department of Physics and Astronomy, Uppsala University, Box 524, 751 20 Uppsala, Sweden} 
\author{Laura Fanfarillo}
\affiliation{Istituto dei Sistemi Complessi (ISC-CNR), Via dei Taurini 19, I-00185 Rome, Italy}
\author{P. J. Hirschfeld}
\affiliation{Department of Physics, University of Florida, Gainesville, Florida 32611, USA}%

\date{\today}

\begin{abstract}
 Since the prediction of a time-reversal symmetry breaking (TRSB) $d+id^\prime$ state in twisted bilayer cuprate superconductors by Can et al. [Nat.Phys. 17,519(2021)], several experiments have attempted to detect this state, yielding conflicting results. At present, it is not clear which differences in samples or experimental conditions might explain these discrepancies.  In this work, we perform a tight-binding lattice model calculation with phenomenological interlayer tunneling, examining  the order parameter as a function of twist angle, interlayer tunneling, doping, and temperature. We observe the TRSB state to be correlated to the position of the Van Hove singularity in the normal state, which changes not only as a function of doping but also the tunneling strength. Two such phases are identified as nominally consistent with in-plane $d+id'$ and $d+is$ order,  but with unexpected transformation properties under  bilayer symmetry operations.  We calculate the Josephson critical current, in particular examining the angle dependence for various tunneling strengths.  Finally, we discuss the existing experiments in the context of our results. 
\end{abstract}

\maketitle

\section{\label{sec:intro} Introduction}
Early theoretical studies of $c$-axis cuprate Josephson junctions \cite{Kuboki1996,Kuboki1998}  predicted that as one twists two two-dimensional(2D) $d$-wave superconductors from $0 \degree$ to $45 \degree$ with respect to each other, the  system should undergo a phase transition from a normal superconductor to a time reversal symmetry breaking (TRSB) superconductor with $d+id^{'}$ order parameter. Due to order parameter mismatch between the two monolayers at $45\degree$, first order pair tunneling processes are expected to vanish. Interest in this phenomenon was revived recently by a more thorough theoretical study by Can et al.~\cite{Can2021} on 2D twisted cuprates, pointing out the expected existence of topological edge currents in the TRSB state. In addition to  Ginzburg Landau calculations, Can et al. employed a continuum model  and  a microscopic lattice model calculation to solve the superconducting(SC) gap equation.  In the lattice model for the parameters of their calculation they predicted the TRSB topological state for  angles ranging from $37\degree$ to $53\degree$.\par
\setlength{\parindent}{20pt}

Since bulk BSCCO-2212 is an extremely 2D system, similar effects are expected in $c$-axis Josephson experiments on twisted crystals, In fact, twisted $c$-axis  tunneling experiments have a long history, but they have tended to report mixed results for the angle dependence of the Josephson critical current \cite{Li1999, Takano2002, Latyshev2004, Lee2021}, and the differences are not generally understood. Recently, Zhao  et al.~\cite{FrankZhao} showed in their cryogenically stacked devices  a strong decrease in the critical current as twist angle was increased, with a small but nonzero residual critical current at $45\degree$. This effect was attributed to second order tunneling processes,  although the magnitude was larger than expected.   Yuan et al.~\cite{PhysRevB.108.L100505} then pointed out that the inhomogeneity in BSCCO could generate an enhanced second order tunneling with magnitude consistent with the measurements.  Other signatures of TRSB behavior reported by Zhao et al. include observing a $\pi$-shifted Josephson Fraunhofer pattern  as well as half-integer  Shapiro steps.

\par
On the other hand, in experiments by Zhu et al.,\cite{Xikunxue} Josephson junctions were made using two ultrathin BSSCO flakes on SiO$_2/$Si twisted with respect to one another at room temperature and annealed at high temperatures. The critical current was found to be only weakly dependent on the twist angle, with no vanishing or strong suppression of the critical current near $45 \degree$, suggesting the absence of the topological state. Possible alternative suggestions for the lack of angle dependence include surface roughness of the substrate on which the exfoliated thin films are placed, variation of tunnel barrier thickness, spontaneous breaking of the fourfold rotational symmetry of the twist junction, and  possible impurities in the tunneling barrier introduced in the fabrication process leading to complex tunneling process through the BiO layer\cite{Xikunxue}.  Zhu et al.\cite{zhu2025manipulatingfractionalshapirosteps} also observed fractional Shapiro step-like features, but pointed out a possible alternative explanation in terms of  vortices.

\par
Cryogenic stacking techniques to fabricate the twist junctions reduce disorder effects but experiments still show  variability: both  Zhao et al.~\cite{FrankZhao} and Martini et al.~\cite{MARTINI} exhibited signatures of TRSB, but Wang et al.~\cite{Wang2023} did not. 
Only recently, Confalone et al.\cite{Poccia2025} suggested that the twisted Josephson junctions should be benchmarked using the critical temperature, the Josephson current density and the junction voltage ($I_\mathrm cR_\mathrm n$). Notably, only the work by Confalone et al.\cite{Poccia2025} reproduced all three parameters previously measured by Zhao et al.
\par
 
\par

Here, we examine possible explanations for the observed experimental variablity in terms of existing theoretical models.    Continuum model calculations  defined in the momentum space lack microscopic detail, but have the advantage of accessing all twist angles continuously and provide a simple overall picture of possible accessible phases. The phase diagram for such a model with fixed filling and tunneling strength predicts the existence of a $d+is$ state for all twist angles, and finds the $d+id'$ state only for twist angles close to $45\degree$~\cite{PhysRevB.105.064501,PixleyVolkov_review}. The $d+is$ states are stable only far below $T_c$, and coexist there with the $d+id'$ state near $45\degree$; however at higher temperatures only the $d+id'$ state is stable\cite{PixleyVolkov_review}.
\par

The details of the phase diagram are also sensitive to the form of the tunneling matrix element between layers.  Most models adopt a simple isotropic interlayer hopping, whose magnitude decays exponentially with distance between the pairs of interlayer sites.
Calculations have been performed in the $t-J$ model where the interlayer tunneling was modeled by taking the orbital symmetry of the copper and oxygen atoms into consideration \cite{Song2022}. In this approximation, it was observed that the state close to $45\degree$ was TRSB but not topological.
\par
The main purpose of this study is to explore the phase diagram of the bilayer system and identify the physical parameters that control the stability of TRSB states. In particular, we aim to understand whether variations in twist angle, filling, or interlayer tunneling can account for the discrepancies between recent critical‑current measurements~\cite{FrankZhao,Xikunxue} and the variability observed in earlier c‑axis twist‑junction experiments \cite{Li1999, Takano2002, Latyshev2004, Lee2021}.
To this end, we employ a microscopic lattice model, which, while computationally demanding, provides a realistic description of the system and precise control over its parameters. We analyze how tuning the microscopic inputs can drive transitions between TRSB and time‑reversal‑symmetric (TRS) superconducting instabilities. We do not include disorder, which can induce states near the Fermi level even in the topological phase~\cite{PhysRevB.110.134514}.
\par
It is instructive to examine the full SC   order parameter, which is a complicated and sometimes counterintuitive object when expressed in terms of sites in the moir\'e unit cell. Here we give explicit procedures to project the full self-consistent solution onto allowed irreducible representations (irreps), and to identify the internal phase between those states which condense. We show that the symmetry of the twisted bilayer is $D_4$, which has consequences for allowed pairing states with symmetry classes distinct from the standard one-plane TRSB states.
\par
Our conclusions for the phase diagram are very roughly consistent with earlier work \cite{PhysRevLett.127.157001}, but present a much richer picture of the order parameter. Some previous lattice calculations, e.g. Ref. \cite{Fidrysiak2023}, simplify this order by minimizing the free energy within an Ansatz $\Delta=\Delta_{1d}+e^{i\phi}\Delta_{d2}$, i.e. assume $d$-wave forms in layers 1 and 2 distinguished only by a complex internal phase.  Instead, we start with interactions leading to nearest-neighbor bond order parameter, solve self-consistently for the order parameter on every bond, and project onto the given local irrep to determine the symmetry.  We further  extend the phase diagrams into larger regions of parameter space, where it is possible to see how different twist junctions may induce states with different properties. 
To help understand why the various TRSB states are stable for different  model parameters, we examine the corresponding Fermi surfaces and sublattice spectral functions. 
 Finally, we calculate the critical current for various angles and tunneling strength to access different parts of the phase diagram and provide a possible explanation for the experimental variability in cuprate twist junctions.

\begin{figure}[tb]
   \centering
    \includegraphics[width=0.8\linewidth]{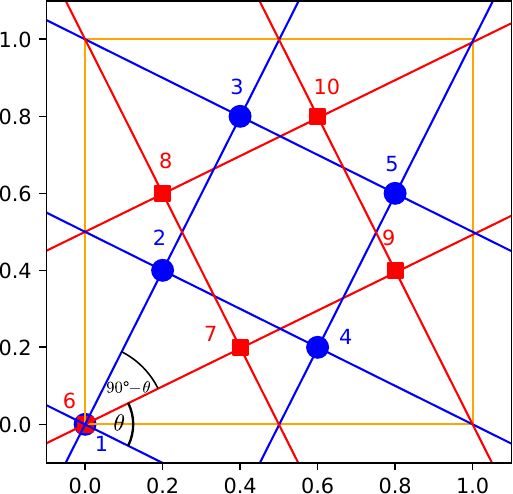}
    \caption{Supercell for $(m,n)=(1,2)$ corresponding to twist angle $\theta =53.13\degree$. In our manuscript we have written $\theta>45\degree$ with the conjugate angle $90\degree-\theta$ which in this case is $36.87\degree$.  Each monolayer has five lattice sites twisted by $\theta/2$ in opposite directions. The two monolayers along with the sites and their labels in each monolayer are represented by  blue and red. }
    \label{supercell}
\end{figure}

\section{Lattice model}
\begin{table*}[tbh]
\centering
\begin{tabular}{|p{1cm}|p{4cm}|p{1cm}|p{1cm}|p{1cm}|p{1cm}|p{1cm}|p{6.5cm}|}
\hline
Irreps & In-plane order parameter & E &  $2C_4(z)$ & $C_2(z)$ & $2C^\prime_2$ & $2C^{\prime\prime}_2$ & basis functions (irrep notation)  \\
\hline
$A_1$ & $s$ wave & +1 & +1 & +1 & +1 & +1 & $x^2+y^2$ $(A_1^1)$, $z^2$ $(A_1^2)$  \\
\hline
$A_2$ & $g$ wave & +1 & +1 & +1 & -1 & -1 & $z$ $(A_2^1)$, $z^3$ $(A_2^2)$, $z(x^2+y^2)$ $(A_2^3)$\\
\hline
$B_1$ &  $d_{x^2-y^2}$ & +1 & -1 & +1 & +1 & -1 & $x^2-y^2$ $(B_1^1)$ , $xyz$ $(B_1^2)$  \\
\hline
$B_2$ &$d_{xy}$ &+1 & -1 & +1 & -1 & +1 & $xy$ $(B_2^1)$, $z(x^2-y^2)$ $(B_2^2)$  \\
\hline
$E$ & $p$ wave & +2 & 0 & -2 & 0 & 0 & $(x,y)$, $(xz,yz)$, $(xz^2,yz^2)$, $(xy^2,x^2y)$, $(x^3+y^3)$  \\
\hline
\end{tabular}
\caption{ Character table for the $D_4$ group. We project our converged order parameter onto the basis functions of each irrep (notation given in brackets)  and the non zero contributions determine the order parameter symmetry.  The in-plane order parameters for the different irreps are the $s$ wave, $d_{x^2-y^2}$ wave, $d_{xy}$ wave etc. We refrain from using these labels to identify states in our calculations because they do not contain the full symmetry of the irrep.  }
\label{CharacterTableD4}
\end{table*}
In lattice model calculations of  moir\'e materials we define the model at commensurate angles so as to maintain periodic boundary conditions. These commensurate angles are defined by parameters $(m,n)$, the twist angle is  $\theta=2\tan^{-1}(m/n)$ where the two square lattices in our bilayer are twisted by $\theta/2$ in opposite directions. The moir\'e unit cell length is $b=\sqrt{m^2+n^2}\times a$ where $a$ is the single-layer lattice constant and has $2(m^2+n^2)$ sites, half of which belong to each monolayer (see Fig.~\ref{supercell}). For the cuprates, $a$ is the Cu-Cu distance in CuO$_2$ layer.

While we perform our calculations for $\theta$ we make use of the $C_4$ symmetry of the square lattices to rewrite $\theta > 45\degree$ as $90\degree-\theta$ such that all the angles in our study range from $0$ to $45\degree$ to observe the gradual changes in the phase space. We have verified this property by looking at the full phase diagram for $0\degree$ [$(m,n)=(0,1)$]  and $90\degree$ [$(m,n)=(1,1)$] and also for  $53.13\degree$ [$(m,n)=(1,2)$] and $36.87\degree$ [$(m,n)=(1,3)$] for certain range of parameters.
\par

We solve a two-layer BCS model with nearest neighbor attraction in each layer,
\begin{equation}
\begin{aligned}
H = & -\sum_{ij,\sigma d} t_{ij} c_{i\sigma d}^\dagger c_{j\sigma d}  
- \mu \sum_{i\sigma d} n_{i\sigma d} \\
& - \sum_{ij,d} V_{ij} n_{i d} n_{j d}  
- \sum_{ij\sigma} g_{ij} (c_{i\sigma 1}^\dagger c_{j\sigma 2} +h.c.)\ ,
\end{aligned}
\label{eq:Hamiltonian}
\end{equation}
where $i,j$ are the site indices, $ c_{i\sigma d}^\dagger$ and $ c_{i\sigma d}$ are electron creation and annihilation operators.  $d=1,2$ is the monolayer index for the bottom and top monolayer,
$t_{ij}$ is the hopping term
with nearest-neighbor hopping $t=0.153$ eV, and next-nearest-neighbor $t^\prime=-0.45t$, consistent with typical cuprate parameters\cite{Can2021}. All energy units throughout this work are in eV. Here $\mu$ is the chemical potential which is fixed self consistently to keep the number of electrons (filling) fixed. The band fillings are denoted by $\nu$ such that the number of electrons per moir\'e unit cell is equal to $2\nu (m^2+n^2)$. $V_{ij}$ is the nearest-neighbor density-density interaction term factorized in the singlet Cooper channel. 
The pairing interaction used here is similar to the spin-fluctuation interaction used in Ref.~\cite{Maier2011} and identical to the one employed by \cite{Joynt1996}.
We have used $V= 0.2$  eV unless otherwise specified. The layers are coupled by an interlayer tunneling term $g_{ij}$ of the form
 \begin{equation}
     g_{ij} = g_0 e^{-\frac{(r_{ij} - c)}{\rho}} \ ,
     \label{eq:tunnel}
 \end{equation}
where $r_{ij}$ is the total distance between sites $i$ and $j$ in different monolayers. $c=a$  is the interlayer distance and $\rho=0.3a$\cite{Can2021} determines the decay  rate of the tunneling. Length scales in this work are denoted in units of $a=1$. For simplicity, we chose  the inter-planar distance equal to the lattice constant which deviates from the real situation in BSSCO bulk crystals.  The precise interlayer distances in the fabricated Josephson junctions are not known; furthermore, larger values of $c$ simply scale the value of $g_{ij}$, such that    the qualitative  features of the phase diagram are unchanged. 
\par
 The Hamiltonian in Eq.~(\ref{eq:Hamiltonian}) is redefined in the moir\'e basis before we perform our calculations. We relabel $i\rightarrow(\alpha,\bf u)$ and $j\rightarrow (\beta,\bf v)$  where, $\alpha,\beta$ are the site labels inside a moir\'e unit cell and $\bf u,\bf v$ are the moir\'e unit cell coordinates. We discuss the details of this redefinition along with our algorithm in Appendix~\ref{algorithm}.  
\par
Within mean field theory, we solve the SC gap equation self consistently,
\begin{equation}
    \Delta_{\alpha\beta}(\mathbf k)=\sum_{\bf k^\prime}V^{\mathbf k \mathbf k^\prime}\langle c_{\alpha \mathbf k^\prime\up} c_{\beta-\mathbf k^\prime\down} \rangle,
\end{equation}
where
\begin{equation}
V^{\mathbf k \mathbf k^\prime}=\sum_{\bf r}Ve^{-i(\mathbf{k}-\mathbf{k}^\prime)\cdot \bf r},
\end{equation}
 $\alpha$ and $\beta$  correspond to site labels in our moir\'e unit cell,  $\bf k, k'$ are momenta in the moir\'e Brillouin zone and $\bf r=\bf u-\bf v$ is the difference between the nearest neighbor moir\'e unit cell coordinates.
Examining the symmetry of this cell with arbitrary (commensurate) twist, together with  the couplings to its neighbors, allows one to conclude that the symmetry group of the twisted lattice bilayer is $D_4$, with character table  shown in Table.~\ref{CharacterTableD4}. 
As a concrete example, the cell for a twist angle of $53.13\degree(36.87\degree)$ is shown in Fig.~\ref{supercell}. Identifying the  symmetry group of the moir\'e lattice then allows us to classify the  order parameter symmetry of all SC states by projecting the converged order parameter onto the $D_4$ irreps.  Using the in-plane basis functions to identify the bilayer order parameters in our calculation does not provide information about the  full symmetry of the irrep and hence we denote the order parameter symmetry with the $D_4$ irrep labels. For example, $A_1 \rightarrow \text{``}s\text{''} $ wave, $B_1\rightarrow  \text{``}d_{x^2-y^2}\text{''} $ wave, $B_2\rightarrow \text{``}d_{xy}\text{''}$ wave.  We describe the procedure in more detail in later sections. 
\begin{figure*}[tbh]
    \centering
    \includegraphics[width=0.95\linewidth]{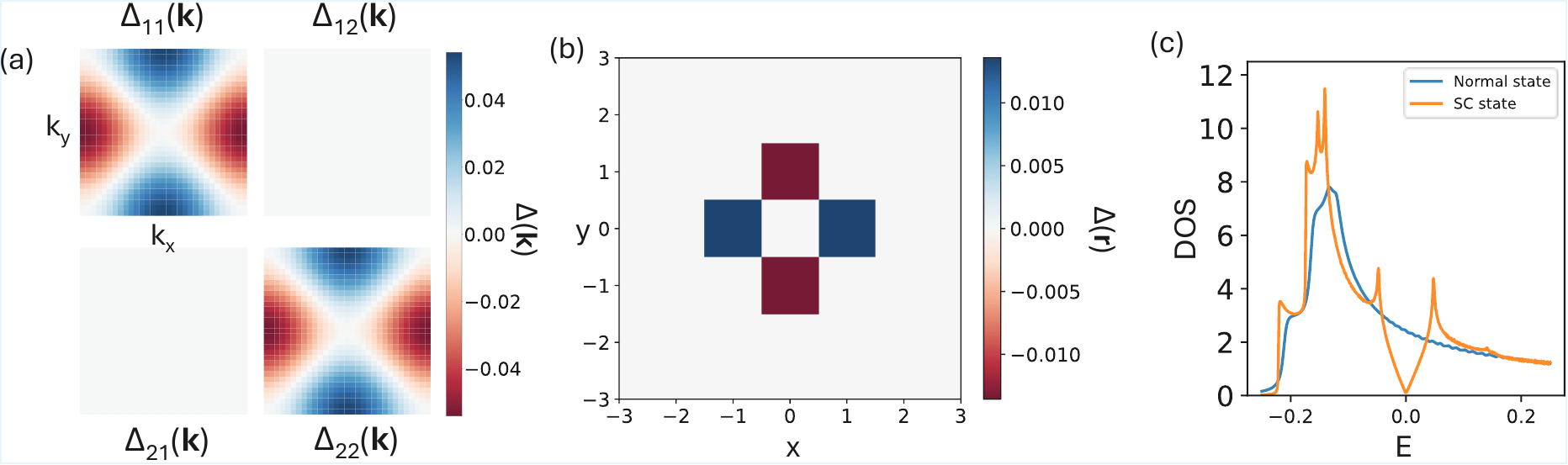}
    \caption{Order parameter and DOS for $\theta=0\degree$ for filling $\nu=0.49$, interaction strength $V=0.2$ and tunneling strength $g_0=0.01$. (a) $B_1$-wave order parameter in momentum space. The $2\times2$ matrix form of the order parameter is because we have 2 sites in our moir\'e unit cell. We clearly see the nodal $d_{x^2-y^2}$ gap along the diagonal elements of the order parameter matrix. (b) $d$-wave order parameter in real space. The order parameter along the $x$ and $y$ direction have opposite signs. (c) DOS displaying the V-shaped DOS which is a signature of $d_{x^2-y^2}$ order parameter. 
    }
    \label{fig_01}
\end{figure*}

For each $\alpha,\beta$, we define the order parameter in the momentum space   spanned by the moir\'e Brillouin zone. To ensure unbiased convergence of all self-consistency equations, we begin with a random complex initial guess for the order parameter. 
We then verify that any imaginary component of the converged real‑space solution is not an artifact of the initial guess by applying a gauge transformation that maximizes the real part. A true TRSB state is therefore one where the imaginary part of the order parameter in real space cannot be removed by such a gauge transformation and stays nonzero at the end of this procedure.

When the gap equations have converged,  depending on the commensurate angle ($m,n$), our order parameter will then be a  $2(m^2+n^2)\times 2(m^2+n^2)$ matrix.
The order parameter $\Delta_{\alpha\beta}(\mathbf k)$ connecting different layers remains zero since Eq.~(\ref{eq:Hamiltonian}) contains only in-plane interactions $V_{ij}$, while  the Fourier transformed pair amplitudes  $\langle c_{\alpha\bf k\uparrow } c_{-\beta \bf k\downarrow } \rangle$ are nonzero.
 Furthermore, since we only have nearest neighbor interactions, we expect non-zero order parameter only where $(\alpha,u)$ and $(\beta,v)$ are nearest neighbors.
\begin{figure*}[tbh]
    \centering
    \includegraphics[width=0.98\linewidth]{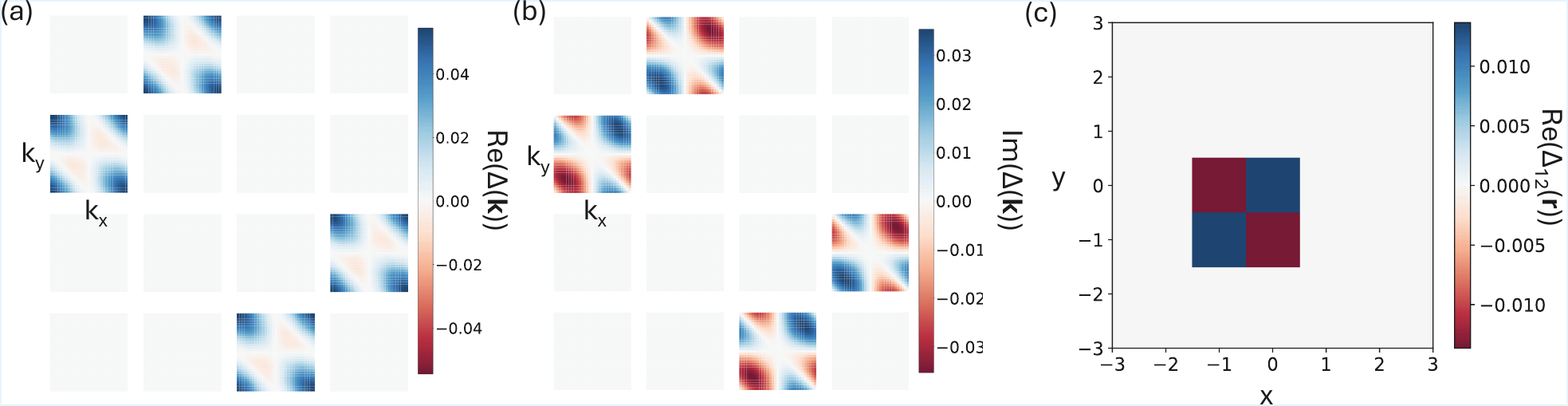}
    \caption{Real and imaginary part of the order parameter for $90\degree$ twist angle for filling $\nu=0.49$, interaction strength $V=0.2$ and tunneling strength $g_0=0.01$. This system has 4 sites in each moir\'e unit cell (Fig. \ref{11lattice}) and hence we have a $4\times4$ order parameter matrix. With two sites in each monolayer the nearest neighbor of site 1 is site 2 and hence we only have off-diagonal elements. (a) Real part of the order parameter and (b) imaginary part of the order parameter have the same form as had been derived in Eq.~(\ref{11 d wave}). (c) The real space order parameter is purely real with non zero entries corresponding to moir\'e unit cell coordinates of the nearest neighbors. All momentum labels are in units of the moir\'e unit cell lattice constant $b$.
    }
    \label{numericalreal11}
\end{figure*}
\par
\begin{figure}[tbh]
    \centering
    \includegraphics[width=0.8\linewidth]{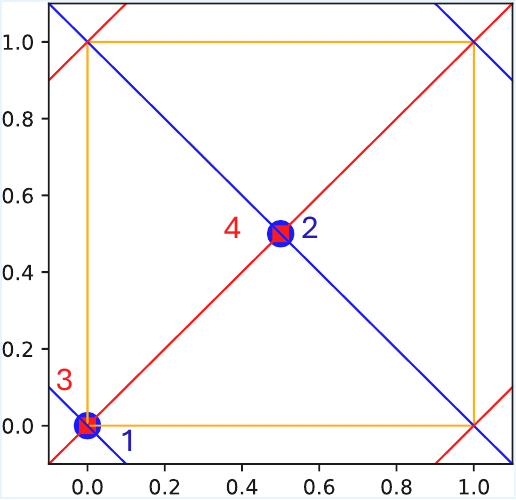}
    \caption{
        Supercell for $(m,n)=(1,1)$ with twist angle $\theta = 90^\circ$. There are four sites in the moir\'e unit cell, 2 in each monolayer represented by color blue and red.}
    \label{11lattice}
\end{figure}
\begin{figure*}[tbh]
    \centering
    \includegraphics[width=\linewidth]{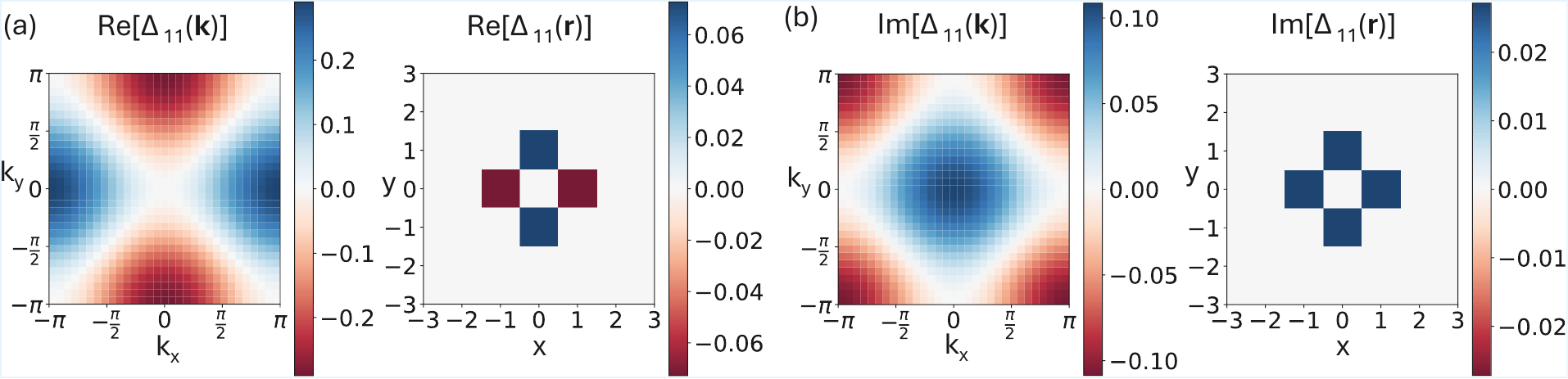}
    \caption{Real and imaginary part of $\Delta_{11}$ both in momentum and real space for $0\degree$ twist angle and filling $\nu=0.4$, tunneling strength $g_0=0.05$ and interaction strength $V=0.5$. (a) Real part of  order parameter in momentum and real space. We observe $d$ wave state. (b) We have a extended $s$ wave order parameter in momentum and real space and a non zero imaginary component in real space  even after performing the gauge transformation.
    }
    \label{d+is zero twist}
\end{figure*}
 
\section{Order parameters for different twist angles.}
\label{TRSBsig}
 Before we scan the full phase space as a function of twist angle ($\theta$), filling ($\nu$) and tunneling strength ($g_0$), we illustrate how the order parameter appears for some special limiting cases. For $0\degree$ twist angle but arbitrary tunneling strength, the model Eq.~(\ref{eq:Hamiltonian}) has been explored earlier in the singlet channel, and is known to exhibit not only $d$-wave pairing, but also $s$ or $s+id$ depending on $g_0$ and $\nu$\cite{Maier2011}.  Since each monolayer is tetragonal, the $90\degree$ twist angle is physically equivalent to $0\degree$; hence, we expect the same state for the same Hamiltonian parameters. 
 The key difference between the representations of the order parameters at $0\degree$ and $90\degree$ arises from the different number of sites in the corresponding moir\'e unit cell, leading to different dimension of the order parameter matrix.  The matrices are physically equivalent but visually different. We first examine a $B_1$($d_{x^2-y^2}$ wave) state for $0\degree$  and the physically equivalent  $90\degree$ twist angle to illustrate the moir\'e representation in some detail, and then consider nontrivial twist cases with these features in mind. 

\subsection{$0\degree$ twist angle}
\par The $0\degree$ twist angle is defined by $(m,n)=(0,1)$, and the order parameter is a $2\times2$ matrix for each momentum point in the moir\'e Brillouin zone. Since at $g_0=0$ and near half-filling we have two decoupled, untwisted layers, we expect for weak coupling of these planes to obtain a stable $B_1$ wave state.  In Fig.~\ref{fig_01}(a), we plot   for  filling $\nu=0.49$,  $g_0=0.01$  the full momentum dependence of the order parameter matrix for each bond $i,j$.  Indeed we  observe a nodal $B_1$ wave order parameter along the diagonals ($\Delta_{11}(\mathbf k)$ and $\Delta_{22}(\mathbf k)$), but since the model includes no interlayer pairing term, we find  $\Delta_{12}(\mathbf k)=\Delta_{21}(\mathbf k)=0$.

Fourier transforming  the converged order parameter from momentum to real space, we observe in Fig.~\ref{fig_01}(b) the expected nearest neighbor $d_{x^2-y^2}$ pairing amplitudes with opposite signs on $x$ and $y$ bonds, as well as the usual V-shaped density of states (DOS) characteristic of  $d$ wave superconductors Fig.~\ref{fig_01}(c).

\subsection{$90\degree$ twist angle}

The $90\degree$ twist angle corresponds to $(m,n)=(1,1)$ with a $4\times4$  order parameter matrix for each momentum point in the moir\'e Brillouin zone. For the same set of parameters as for $0\degree$, the order parameter converges to a $B_1$ wave order parameter, shown in Fig.~\ref{numericalreal11}, which appears rather different from the $0\degree$ case, although it must be physically equivalent.
We have 2 sites per monolayer in our moir\'e unit cell  (Fig.~\ref{11lattice}), such that the analytical solution for the order parameter between sites 1 and 2  for our case of a nearest neighbor pairing interaction  is given by 
\begin{align}
    \Delta_{12}(\bf k)&= \mathop{\Delta_{12}}[(1+\cos(k_x+k_y)\notag
    -\cos(k_x)-\cos(k_y)\notag)\\
	&+i(\sin(k_x+k_y)-\sin(k_y)-\sin(k_x))].
\label{11 d wave}
\end{align}

The  non-zero imaginary part of the order parameter is due to working with the Hamiltonian in the  moir\'e unit cell basis.  We have performed the full derivation in the Appendix~\ref{90derive}.

 As seen in Fig.~\ref{11lattice},  the nearest neighbor of site 1  is site 2 in each moir\'e unit cell.   The moir\'e unit cells with coordinates
 $(0,-1),(-1,0)$ and $(-1,-1)$ (in units of the moir\'e lattice constant $b$)  with respect to the $(0,0)$ cell shown in Fig. \ref{11lattice} enter the Fourier transform, but are not symmetric about (0,0),
 and hence  we obtain a non zero imaginary part of the order parameter.
\par
The results after the self consistency are presented in Fig.~\ref{numericalreal11}. The real part of the order parameter in Fig.~\ref{numericalreal11}(a) is apparently different from the $B_1$ wave order parameter we obtained in Fig.~\ref{fig_01}(a), but follows the analytical form we presented in Eq.~(\ref{11 d wave}) and plotted in Fig.~\ref{90 angle d wave analytical} in the appendix. $\Delta_{12}(\mathbf{k})=\Delta^*_{21}(\mathbf{k})=\Delta_{34}(\mathbf{k})=\Delta^*_{43}(\mathbf{k})$ are the non zero entries because 1, 2 and 3, 4 are nearest neighbors of each other and lie in the same monolayer; all  other entries are zero. The imaginary part of $\Delta_{12}(\bf k)$,  shown in Fig.~\ref{numericalreal11}(b), is not due to having a complex initial guess and cannot be gauged away. 

On performing the Fourier transform, we obtain the real part of the order parameter in real space (Fig.~\ref{numericalreal11}(c)) which seems to be different from Fig.~\ref{fig_01}(b) but is equivalent.  This can be understood by realizing that the non zero entries correspond to the nearest neighbor moir\'e unit cell coordinates as noted above.
The imaginary part of the order parameter in real space is 0 and hence we have a pure $B_1$ wave solution as expected.

\begin{figure}[tb]
    \centering
    \includegraphics[width=0.85\linewidth]{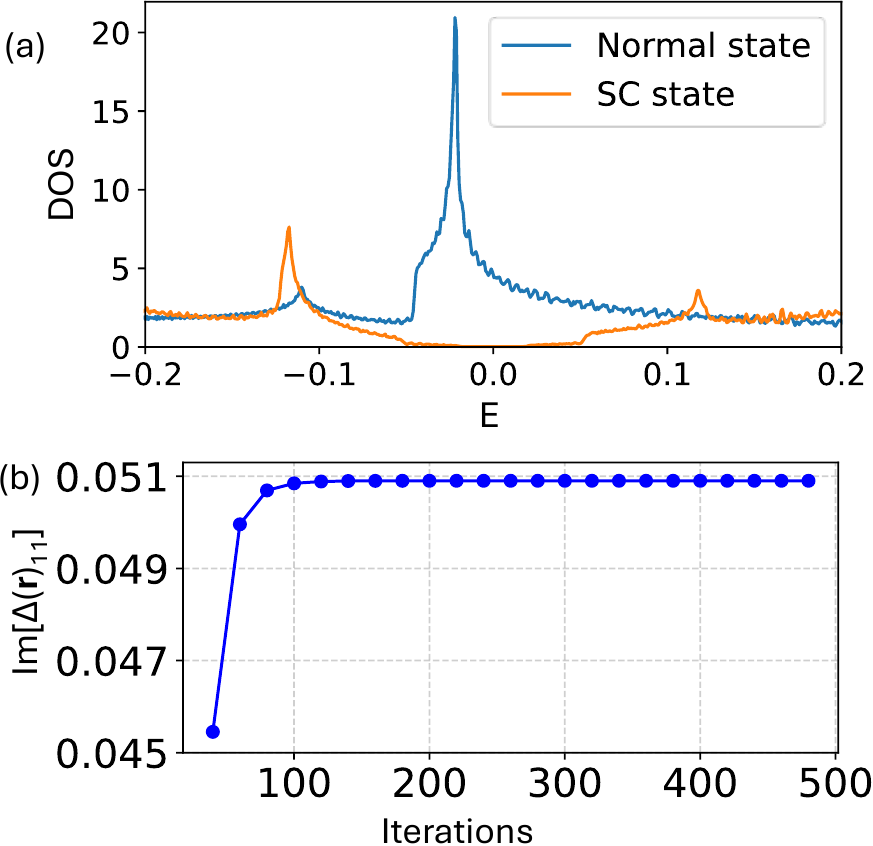}
    \caption{ Signatures of time reversal symmetry breaking for $\theta=0\degree$twist angle case discussed in Fig.~\ref{d+is zero twist}. (a) Gapped DOS  observed in the SC state along with the coherence peaks and Van Hove singularity peaks. (b) Imaginary part of the order parameter in real space for the 11 matrix element vs no of iterations. The convergence of this quantity to a finite non zero value indicates the presence of a TRSB state.}
    \label{d+is zero twist signature}
\end{figure} 

\subsection{Observing time reversal symmetry breaking states for $0\degree$  twist angle.}

Thus far, we have discussed a $B_1$ wave solution for $0\degree$ twist and weak interlayer tunneling. As we vary the interlayer tunneling $g_0$ and the filling $\nu$  we observe TRSB states, e.g. at   $\nu=0.4$, $g_0=0.05$ and interaction $V=0.5$ shown in Fig.~\ref{d+is zero twist}.  We have chosen a larger value of $V$ to obtain a larger SC gap in order to observe and benchmark features of the TRSB states. We use more realistic values for our phase space scan. The real part of the order parameter in momentum and real space is $B_1$ wave state as shown in Fig.~\ref{d+is zero twist}(a);  the imaginary part of the order parameter in momentum space has  $A_1$-wave behavior as shown in Fig.~\ref{d+is zero twist}(b). On performing Fourier transform, we find a non-zero imaginary part of the order parameter in real space which cannot be gauged away. Thus we obtain a $B_1+iA_1$ wave state. This state at nonzero tunneling but zero twist angle is related to the well-known $d_{x^2-y^2}+is$ state found in a somewhat different  filling range at $g_0 \rightarrow 0 $\cite{Maier2011}.
\par

A key difference between a TRS $B_1$ wave state and a fully gapped TRSB $B_1+iA_1$ or $B_1+iB_2$ state is the DOS, since any nodes are lifted by the  out-of-phase component. We plotted the DOS of the normal and SC state in Fig.~\ref{d+is zero twist signature}(a).  Compared to the normal state DOS we can see the gap opening along with the coherence peaks and shifted Van Hove peaks.
Calculating the DOS with sufficient resolution to distinguish TRSB states from TRS states is computationally challenging since the moir\'e unit cell grows as we approach the $45\degree$ twist. Hence, we examined signatures of TRSB that could be obtained directly from the order parameter.

We investigate the imaginary part of the real space order parameter to differentiate between TRS and TRSB states. For $\theta=90\degree$  we discussed a $B_1$ wave state with complex momentum space order parameter which after the Fourier  and gauge transform  was purely real in real space. 
In contrast, in Fig.~\ref{d+is zero twist signature}(b) we exhibit the imaginary part of the real space order parameter as a function of iterations and its convergence to a nonzero value after gauge transformation,  a signature of a TRSB state.
\par
  \begin{figure*}[tb]
    \centering
    \includegraphics[width=0.85\linewidth]{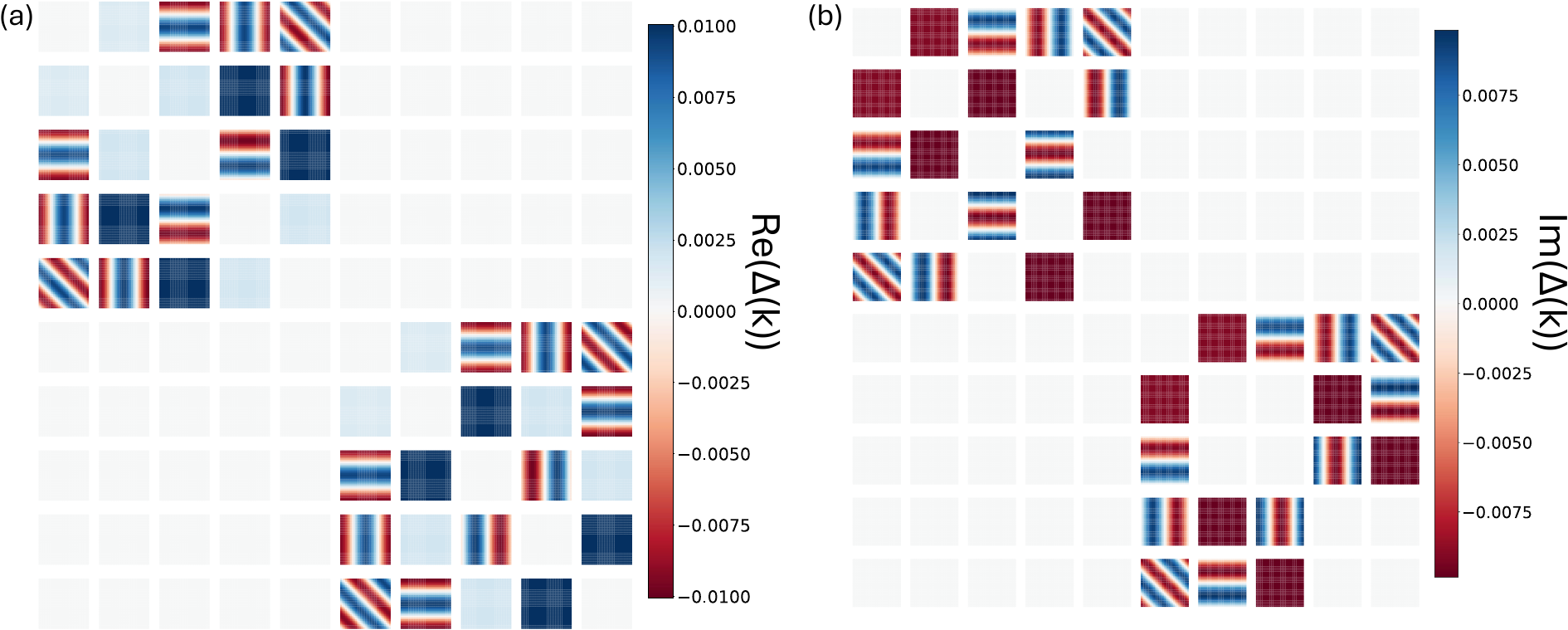}
    \caption{Order parameter for $36.87\degree$ twist angle, filling $\nu=0.4$, tunneling strength $g_0=0.12$, interaction strength $V=0.2$. There are 10 sites in our moir\'e unit cell 
    (Fig. \ref{supercell}) and thus we have a $10\times10$ matrix. (a) Real part of the order parameter. (b) Imaginary part of the order parameter.}
    \label{TRSB OP 53.13 real}
\end{figure*}
The continuum model calculation predicted regions of $d+is$ order at low temperatures for small twist angles, and coexisting $d+is$ and $d+id$ phases for angles close to $45\degree$ \cite{PhysRevB.105.064501}. Neither a gapped DOS nor  a finite imaginary part of the order parameter in real space provide us with any information about the order parameter symmetry here. Looking at the order parameter momentum dependence ($\Delta(\mathbf{k})$) to identify the symmetry is also challenging for $\theta$ close to $45\degree$ because of the complicated order parameter matrix.
However, the exact symmetry of the order parameter is directly obtainable by Fourier transforming $\Delta_{\alpha\beta}(\mathbf{k})$ to $\Delta_{\alpha\beta}(\mathbf{r})$ and projecting it onto each irrep basis function of the $D_4$ group. We use the coordinates of the center of each bond during projection, translating bonds which lie outside the moir\'e unit cell by the moir\'e unit cell lattice vectors into the elementary cell. The nonzero components after projection identify the order parameter symmetry. We only project the order parameter onto a few low order basis functions for each irrep and not all harmonics, hence the magnitude of the order parameter for each irrep ($\Delta^{\Gamma}$) is not properly normalized and is reported below in arbitrary units (a.u).

 \begin{figure}[tb]
    \centering
    \includegraphics[width=0.85\linewidth]{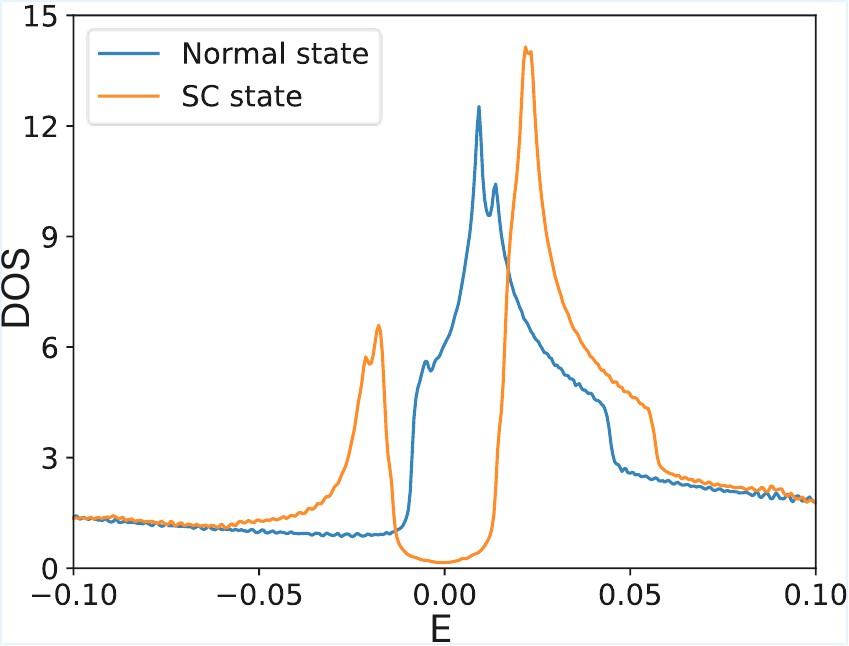}
    \caption{DOS for $\theta=53.13\degree$, filling $\nu=0.4$, tunneling strength $g_0=0.12$ and interaction strength $V=0.2$ case along with the normal state DOS. We observe the gap opening which is a signature of the TRSB state. }
    \label{21DOS}
\end{figure}

\begin{figure*}[tb]

 \includegraphics[width=\linewidth]{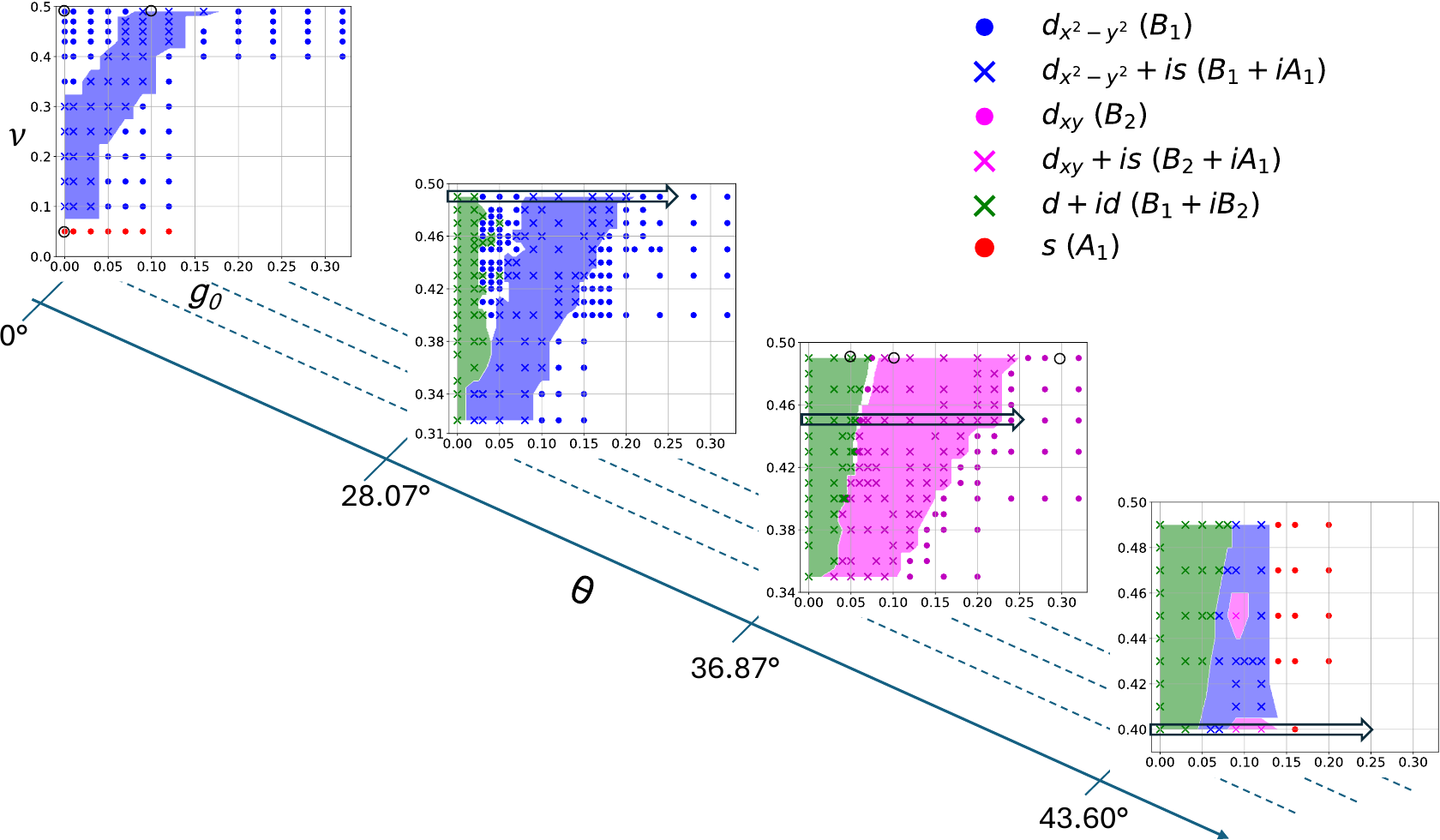}
 
         \caption{Phase diagram for twisted bilayer cuprate as a function of twist angle ($\theta$), filling ($\nu$) and tunneling strength ($g_0$) for fixed interaction strength $V$=0.2. Each panel represents the phase diagram for a particular twist angle as a function of filling and tunneling strength. The blue dots represent TRS $B_1$ wave states, the blue crosses are for $B_1+iA_1$ states and the blue regions mark the phase boundary for this state. Magenta dots are for $B_2$ states, magenta crosses represent the $B_2+iA_1$ state with the magenta colored region being the phase boundary for this particular TRSB state. The $B_1+iB_2$ state is represented by green crosses and the phase boundary by the green region. Finally, the pure $s$ wave states are represented by red dots. The black circles in $\theta=0$ and $\theta=36.86$ represent the points in the phase diagram for which we have investigated the sublattice spectral functions and Fermi surface in Fig.~\ref{Spectral fucntions}. The arrows in the phase diagram of $\theta=28.08\degree$, $\theta=36.87\degree$ and $\theta=43.60\degree$ are the regions for which we have examined the evolution of each irrep in Fig.~\ref{Opvsgo}.}
         \label{Phase diagram}
 \end{figure*}

\subsection{$36.87\degree$ twist angle.}
The $36.87\degree$ twist angle is the conjugate of $53.13\degree$  defined by  $(m,n)=(1,2)$. The order parameter is a $10\times10$ matrix for each $\mathbf{k}$ point. Using the parameters $\nu=0.4$ and $g_0=0.12$, we converged to a $B_1+iA_1$ state which is shown in Fig.~\ref{TRSB OP 53.13 real}

The momentum dependence of each matrix element $\Delta_{ij}(\mathbf{k})$ is dependent on the moir\'e unit cell coordinates of the nearest neighbor. For example, $\Delta_{13}(\mathbf{k})$ has a momentum dependence along the $y$ direction because the nearest neigbour bond between site $1$ and site $3$ connects moir\'e unit cells with coordinates $(0,0)$ and $(0,-1)$. Similarly for $\Delta_{14}$ the bond is along $(-1,0)$  and for $\Delta_{15}$ the bond is along $(-1,-1)$ and hence the corresponding momentum dependence is along $x$ axis and the diagonal respectively.
\par

To determine whether the state is TRSB or not, we evaluated the DOS and observed the gap shown in Fig.~\ref{21DOS}. The occurrence of a gapped $B_1+iA_1$ state for twist angles close to $45\degree$ can therefore provide an explanation for significant Josephson currents for twist angles close to $45\degree$ \cite{Xikunxue}. Since other gapped states are also possible within the same model, we explore other parameter ranges to ascertain which gapped ground states are plausible.

\begin{figure*}[tb]

\includegraphics[width=\linewidth]{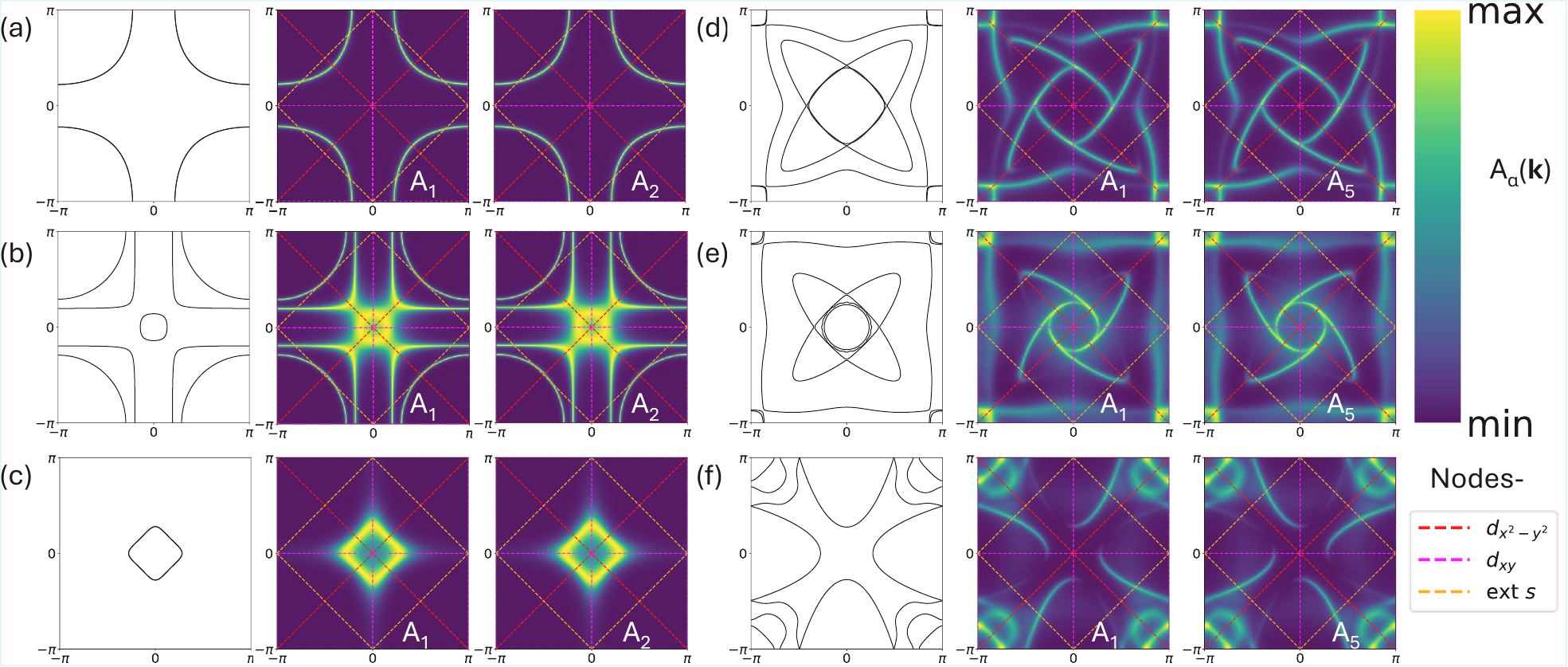}
\caption{Fermi surface and sublattice spectral function ($A_{\alpha}(\mathbf{k})$) for various parameters in the phase diagram. The overlap  of the spectral function with the antinodes of an order parameter symmetry reduces the condensation energy hence favoring a particular state.  The $\Gamma$ point in located at $(0,0)$, the $M$ points are located along the corners $(\pi,\pi), (-\pi,\pi), (\pi,-\pi)$ and $(-\pi,-\pi)$, and the X and Y points are located at $(\pi,0),(-\pi,0)$ and $(0,\pi),(0,-\pi)$ respectively. The observed phases in the phase diagram can be understood by looking at the sublattice spectral functions for certain cases. (a) $\theta=0\degree$, $g_0=0$, $\nu=0.49$, $B_1$ wave. (b) $\theta=0\degree$, $g_0=0.1$, $\nu=0.49$, $B_1+iA_1$ wave. (c) $\theta=0\degree$, $g_0=0$, $\nu=0.05$, $A_1$ wave.(d) $\theta=36.87\degree$, $g_0=0.05$, $\nu=0.49$, $B_1+iB_2$ wave. (e) $\theta=36.87\degree$, $g_0=0.1$, $\nu=0.49$, $B_2+iA_1$. (f) $\theta=36.87\degree$,    $g_0=0.3$, $\nu=0.49$, $B_2$ wave.}
         \label{Spectral fucntions}
    
 \end{figure*}

\section{Phase Diagram}
\begin{figure}[tb]

 \includegraphics[width=0.85\linewidth]{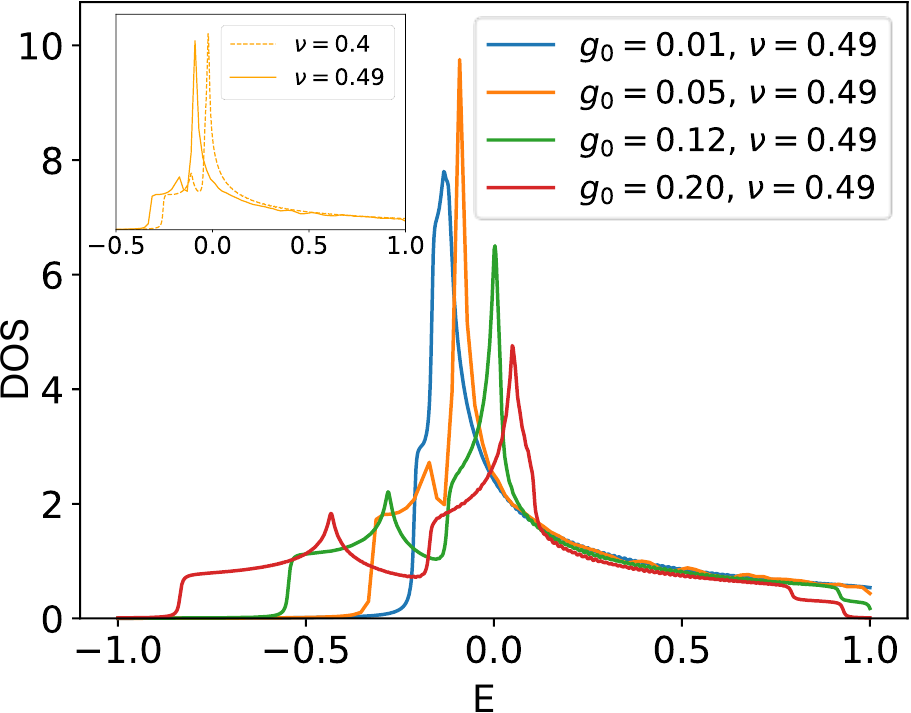}

         \caption{Normal state DOS vs E for different $g_0$ \ for $\theta=0$. The Van Hove singularity splits and shifts as we increase $g_0$ and we observe $B_1+iA_1$ states in region where the Van Hove singularities lie in proximity of the chemical potential. The insert shows a shift of the Van Hove singularities closer to the chemical potential as we decrease filling for $g_0=0.05$. }
         \label{Dos Compare}
    
 \end{figure}

\par

 The observation of a $B_1+iA_1$ state at $36.87\degree$ where $B_1+iB_2$ was predicted earlier\cite{Can2021} (at smaller $g_0$) already hints at a transition from one TRSB state to another.  Clearly we would like to identify all possible TRSB phases (identified in Fig. \ref{Phase diagram} by shaded regions), and any other fully gapped phases.   In this section, we therefore discuss the $\nu-g_0$ phase diagram, shown in Fig. \ref{Phase diagram}, for various commensurate angles $\theta$ at temperature $T=10^{-4}\ll\Delta$.  We first describe  the phases observed at the various angles considered, and subsequently advance some ideas why these  phases exist and why certain transitions occur.
\par  
For the model given by Eq.~(\ref{eq:Hamiltonian}) on a single layer square lattice,  as one dopes away from half filling to lower fillings, the  $d$ wave ground state gives way to a $d+is$ phase and eventually to  a $s$ wave state\cite{Joynt1996,Breio2022}. For $\theta=0\degree$ and $g_0=0$, the two monolayers in our model are decoupled and hence we expect to see similar feature in our phase diagram. As seen in Fig. \ref{Phase diagram},  a pure $B_1$ wave state spans the filling range $\nu\in[0.49,0.35]$, $B_1+iA_1$ state spans $\nu\in[0.35,0.1]$ and the $A_1$ wave state is stable for $\nu\in[0.1,0.05]$.  Thus the search for TRSB states begins already with the decoupled untwisted bilayer $B_1+iA_1$ state, which is  also stabilized at fillings closer to half-filling as $g_0$ increases.  

\par
At $28.08 \degree$ twist angle, we perform the phase space scan for $\nu\in[0.3,0.49]$. On increasing $g_0$ we move from a green region of $B_1+iB_2$ wave states to a blue region of $B_1+iA_1$ wave states separated by a narrow region of $B_1$ wave states which are also present for large $g_0$. The phase diagram for $36.87\degree$ is qualitatively similar to $28.08\degree$. We performed the scan for $\nu\in[0.35,0.49]$. We observe that the area of the $B_1+iB_2$ state is significantly larger and persists up to higher tunneling strengths and the region of TRS $B_2$ wave state separating $B_1+iB_2$ and $B_2+iA_1$ phase is significantly reduced. The closest angle to $45\degree$ for which we performed calculations was $43.60\degree$ for $\nu\in[0.4-0.49]$. The area of $B_1+iB_2$ state has further increased. The region of $B_2$ wave state separating the $B_1+iB_2$ and $B_2+iA_1$ has vanished and instead we have coexistence region where both $B_1+iB_2$ and $B_2+iA_1$ state exist.  The $g_0=0$ limit of the calculation physically means two decoupled layers. The identification of TRSB states in that limit is due to the order parameter symmetries being defined relative to the moir\'e unit cell for each angle and thus at non-zero twist angles we have finite $B_1$ and $B_2$ contributions with a complex internal phase generated due to a random initial guess. We have explicitly shown the evolution of the projected order parameters for various $\theta$ in Appendix~\ref{Opevol}.

We note that the various possible ground states found are expected to be very close in energy as the Fermi surface becomes more complicated at high twist angles.  Performing self consistent microscopic calculations can become computationally expensive as we approach twist angles closer to $45\degree$, and with our current momentum grids ($31\times31$ for $\theta=36.87\degree$ and $13\times13$ for $\theta=43.6\degree$), eigenvalue crossings with $g_0$ or $\nu$ are not always resolved accurately.
One possible such example is  for $\theta=28.08\degree$, we observed some isolated $B_1+iB_2$ solutions for  $\nu=0.47$, $g_0=0.05$ immersed in regions where $B_1$ or $B_1+iB_2$ states appear to be dominant.

Understanding the stability of the various phases found requires a knowledge of the position of the Van Hove singularity in the model relative to the Fermi level, as well as the strength of the pairing interaction evaluated on nested regions of the Fermi surface.  We begin at  $\theta=0\degree$ by plotting in Fig.~\ref{Dos Compare} the normal state DOS as a function of $g_0$ and $\nu$.
The solid lines in the figure represent DOS for $\nu=0.49$ for various $g_0$. Increasing $g_0$ leads to splitting of the Van Hove singularity and for a range of values it lies in proximity to the chemical potential coinciding with the observed TRSB $B_1+iA_1$ states, eventually moving away. For $\nu=0.49$, the range is $g_0\approx[0.07,0.16]$. The insert in Fig.~\ref{Dos Compare} displays the shift of the Van Hove singularity as we change filling. We observe that for the same $g_0=0.05$ the Van Hove singularity  for $\nu=0.4$ is shifted right and hence closer to the chemical potential as compared to $\nu=0.49$ indicating that we need a smaller $g_0$ to bring the Van Hove singularity close to the chemical potential as we move away from half filling. This explains why we observe TRSB at smaller $g_0$ as we decrease filling in Fig.~\ref{Phase diagram}.   
\par
To understand the different phases in our phase diagram we investigate the normal state sublattice spectral function, defined as diagonal components of the full spectral function matrix
\begin{align}
    A_{\alpha}(\mathbf{k},\omega)=-\dfrac{1}{\pi}\mathop{\text{Im}}[(\omega+i\eta)\mathbbm{1}-H_0(\mathbf k)]^{-1}_{\alpha\alpha} ,
\end{align}
where $H_0$ is the non interacting part of the Hamiltonian as derived in Eq.~(\ref{eq:non interacting Hamiltonian}), $\mathbbm{1}$ is an identity matrix and   $\alpha$ is the moir\'e site index. For the Fermi surface $\omega=0$ and hence we examine $A_{\alpha}(\bf k)$ to obtain some key insights  about the evolution of the order parameter across the phase diagram. The presence of high spectral weight in the Fermi surface anti-nodal regions  of a given order parameter symmetry favors the state by lowering the condensation energy.
We first consider various cases for the untwisted bilayer $\theta=0\degree$ to understand the different phases. Fig.~\ref{Spectral fucntions}(a) is simply the Fermi surface for  $\nu=0.49$ and $g_0=0$,  together with plots of the spectral functions of the two sites in the moir\'e unit cell, identical in this case.  The sheets at the $M$ point  have spectral weights near X and Y points which are the antinodes of the $B_1$ order parameter, thus stabilizing this state. While the $M$ point is also an antinode for the $A_1$ phase, this particular spectral function is plotted very close to a phase boundary where details of the Fermi surface and the evolution of the spectral function for energies of the order of the SC gap determine the result of the competition between $B_1$ and $A_1$. Fig~\ref{Spectral fucntions}(b) is plotted for  $\nu=0.49$ and $g_0=0.1$, where the Fermi surface is already considerably more complicated, including a small pocket around $\Gamma$.  As seen in the spectral function plots, the $B_1+iA_1$ wave state is stabilized by the combination of spectral weight close to the $X$ and $Y$ points and the $\Gamma$  point in the normal state, i.e.  regions near the antinodes of  $B_1$ and $A_1$ irrep. There are no contributions at the M point which is an antinode of both $B_2$ and $A_1$ wave.  A simple confirmation is provided in Fig~\ref{Spectral fucntions}(c),  plotted for small filling $\nu=0.05$ and $g_0=0$ with only one pocket around the $\Gamma$ point which stabilizes   $A_1$ wave (extended-$s$) state, the only state with an antinode there. 

In Fig. \ref{Spectral fucntions}(d)-(f), we now  investigate similar features for $\theta=36.87\degree$ to obtain an idea about the transitions between the different phases at nontrivial twist. Fig.~\ref{Spectral fucntions}(d) shows the Fermi surface and sublattice spectral weight for  $\nu=0.49$ and $g_0=0.05$. The presence of significant $A_1(\bf k)$ and $A_5(\bf k)$ spectral weight in the antinodal region of $B_1$ and $B_2$ and its absence at the $\Gamma $ point which is the other  antinodal region of $A_1$ wave favors a $B_1+iB_2$ state, as observed in Fig. \ref{Phase diagram}. 
Fig.~\ref{Spectral fucntions}(e) is plotted for $\nu=0.49$ and $g_0=0.1$ where the spectral function on the $\Gamma$ pocket is much larger, thus leading to a $B_2+iA_1$ state. Fig.~\ref{Spectral fucntions}(f) is for  $\nu=0.49$ and $g_0=0.3$ and the only regions with significant spectral weights lie along the $M$ point but since we have no contribution at $\Gamma$ point  $B_2$ order parameter is stabilized.
\par
\section{Temperature dependence of the order parameter}
The study of the phase diagram and its properties presented thus far have all been done at very low temperatures, where we have shown that  multiple irreps compete with each other.  At nonzero temperatures, all states corresponding to a given irrep must have the same critical temperature, but different irreps $\Gamma$ will condense generally at different $T_c^\Gamma$.  This means that a TRSB state will generally evolve into a TRS state with a single component as the temperature is raised.

\par
\begin{figure}[tb!]
    
 \includegraphics[width=0.8\linewidth]{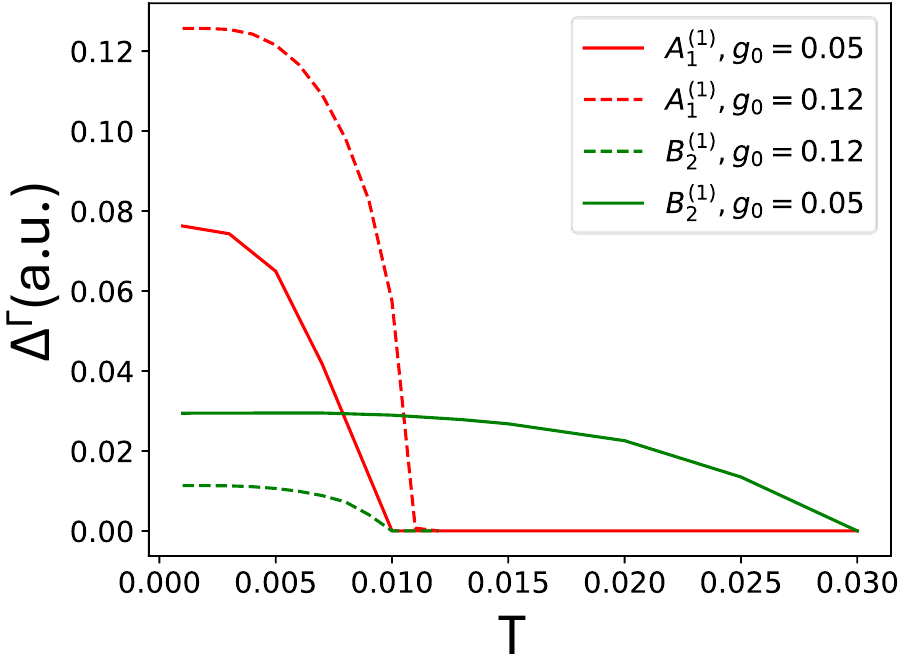}

         \caption{Temperature dependence of different order parameter irreps for $53.13\degree$ twist for $g_0=0.05$ and $g_0=0.12$.}  
         \label{op_vs_T_0_twist}
    
 \end{figure}

Without attempting to present a full phase diagram vs. $T$, we illustrate this phenomenon  by examining the evolution of the different irreps of the order parameter $\Delta^{\Gamma}$ as a function of temperature. In Fig.~\ref{op_vs_T_0_twist} we choose two $B_2+iA_1$ states for $\theta=36.87\degree$, $\nu=0.4$ and $g_0=0.05,0.12$ located on opposite sides of the phase boundary. For $g_0=0.05$ the $A_1$ wave  vanishes first as we increase temperature and the $B_2$ wave persists upto highest temperature thus determining the $T_c$. For $g_0=0.12$ the $A_1$ wave  persists up to a larger $T_c$ whereas the $B_2$ wave irrep vanishes, showing that as a function of temperature  either $A_1$ or {\red $B_2$} wave is stable, depending on Hamiltonian parameters.
 
\section{Josephson Critical Current}
One of the possible explanations for the apparently inconsistent observations of dramatically suppressed  Josephson critical  current ($J_c$) \cite{FrankZhao} and significant $J_c$ \cite{Xikunxue} in twisted BSSCO systems at $45\degree$ twist angle could be different SC states being stabilized for the two experimental setups as established by the competing phases in the phase diagram. Understanding how  $J_c$ varies in these different states could provide us with some insights to the observed differences.
\par
For a given set of model parameters we start with the fully converged order parameter; then, to calculate $J_c$ we change the phase in one of the monolayers by $\phi$ which we then vary to plot a  $J$ vs $\phi$ curve, choosing the maximum value of $J$ as $J_c$. The expression for the same is 
 \begin{equation}
    J(\phi)=-\dfrac{ie}{\hbar}\sum_{(\alpha,\bf u),(\beta,\bf v)\sigma}g_{\alpha\beta}^{\bf u \bf v}\langle  c^{\bf u\dagger}_{\alpha\sigma}c_{\beta\sigma}^{\bf v}-c^{\bf v\dagger}_{\beta\sigma}c_{\alpha\sigma}^{\bf u}\rangle_{\phi},
    \label{eq:Jcurrent}
 \end{equation}
where sites $(\alpha,\bf u),(\beta,\bf v)$ belong to different monolayers and $\langle  c^{\bf u\dagger}_{\alpha\sigma}c_{\beta\sigma}^{\bf v}-c^{\bf v\dagger}_{\beta\sigma}c_{\alpha\sigma}^{\bf u}\rangle_{\phi}$ is obtained after solving the Hamiltonian with the phase changed order parameter to obtain the new creation and annihilation operators in terms of the quasiparticle states.
We define the dimensionless quantity $j_c=(J_c\hbar a^2)/(eg_0A)$ as the normalized $J_c$.  In this definition, we have explicitly divided out the trivial  $g_0$ scaling. $A$ is the area of the moir\'e unit cell for a commensurate twist angle $\theta$. 
$j_c$ is suitable to better understand the effect of the order parameter symmetry on  the measured $J_c$.   We note that for numerical reasons, the order parameter used in the calculation of Eq.~(\ref{eq:Jcurrent}) is not calculated self-consistently in the presence of the current, as has been also the case in previous calculations \cite{PhysRevB.105.064501}.  We believe this does not affect the qualitative results presented below.

\begin{figure}[tb!]
\includegraphics[width=\linewidth]{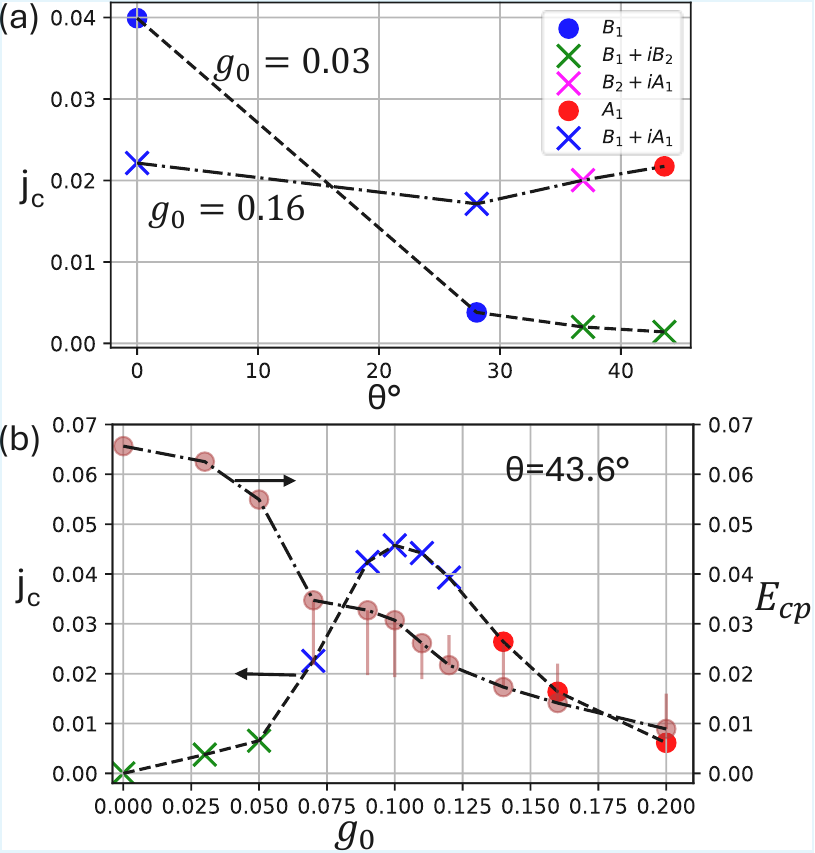}
\caption{Normalized Josephson critical current $j_c=(J_c\hbar a^2)/(eg_0A)$ as the normalized Josephson critical current dependence on Hamiltonian parameters. (a) $j_c$ vs $\theta$ for $\nu=0.49$, $g_0=0.03$ and $\nu=0.49$, $g_0=0.16$. (b) $j_c$ vs $g_0$ and $E_{cp}$ vs $g_0$ for $\nu=0.43$ and $\theta=43.6\degree$. The brown circles  are estimate for coherence peak energies ($E_{cp})$. Some values have an error bar because the coherence peaks are not easily decipherable due to the proximity of the Van Hove peak to the chemical potential. The arrows indicate the y-axis associated for each plot}
\label{Josephson_critical_current}
\end{figure}

As seen in Fig.~\ref{Josephson_critical_current}(a), for a fixed $g_0=0.03$ and $\nu=0.49$, $j_c$ monotonically  decreases by two orders of magnitude from $\theta=0\degree$   as $\theta$ increases to $\theta=43.6\degree$. This result is consistent with the experimental observation by Zhao et al.~\cite{FrankZhao} in twisted BSCCO crystals. For higher $g_0$, we observe very little angle dependence of $j_c$, which is consistent with observations reported by Zhu et al. for twisted films annealed at high temperature~\cite{Xikunxue}.  Thus the discrepancy between the two experiments could plausibly be explained by postulating that the essential difference between the two is a larger effective interlayer coupling in the latter.   We speculate that this may be due to roughness of the surfaces in the Zhu \textit{et.al.} experiement, but there is no  reported experimental evidence for this.  Note that if this explanation is correct, it suggests that the Zhu \textit{et.al.} system may realize completely different ground states than the anticipated $B_1+iB_2$, to wit the TRSB $B_1+iA_1$ or even TRS $A_1$ states.
 \par
To verify how competing order parameters change  $j_c$ for a fixed twist angle we plot $j_c$ vs $g_0$ for $\theta=43.6\degree$, $\nu=0.43$ in Fig~\ref{Josephson_critical_current}(b). We observe that by increasing $g_0$ we move from a $B_1+iB_2$ phase with very small $j_c$ to a $B_1+iA_1$ phase which is a TRSB state but with a larger $j_c$ and finally to the TRS $A_1$ phase. We observe that $j_c$ increases up to $g_0=0.10$ and then decreases.  

\par
To understand the dependence of the critical current on the tunneling energy $g_0$, we plot in the same graph the coherence peak energies ($E_{cp}$) as a proxy  for the overall magnitude of the SC order parameter. Note that in  some  cases the Van Hove peaks lie close to the chemical potential and the coherence peaks are not easily distinguishable, which we indicate with error bars for certain values of $g_0$. We see a monotonic decrease of the order parameter magnitude as $g_0$ is increased, explaining the decrease of $j_c$ in the large tunneling regime.  We believe this is due to deviation from optimal conditions for $d$ wave pairing as $g_0$ is increased.
For $g_0\rightarrow 0$, the order parameter saturates to its value in each isolated layer, while $j_c$ decreases. This behavior can be explained from a perturbative argument how the expectation value $\langle  c^{u\dagger}_{\alpha\sigma}c_{\beta\sigma}^{v}-c^{v\dagger}_{\beta\sigma}c_{\alpha\sigma}^{u}\rangle_{\phi}$ scales as a function of $g_0$. For small $g_0$, the self-consistently obtained order parameter and chemical potential are unaffected, so the onset of the expectation value comes from the scaling of the components of the eigenvectors. The BdG Hamiltonian acquires off-diagonal elements proportional to $g_0$ that couple arbitrary eigenstates with energy $E_1$ and $E_2$. We now examine how the eigenvectors of the following $2\times2$ matrix scales with $g_0$,
\begin{align}
\begin{bmatrix}
E_1 & g_0 \\
g_0 & E_2 
\end{bmatrix}=\frac{E_1+E_2}{2} +\frac{E_1-E_2}{2}\sigma_z+g_0\sigma_x \,.
\end{align}
Calculating the unitary transformation that diagonalizes this matrix, we can express the $c$ operators in the band basis to calculate the expectation value in Eq.~(\ref{eq:Jcurrent}). For small $g_0$, we find that the components of the eigenvectors scale linearly with $g_0$, and therefore so does the expectation value.
Hence for small $g_0$, $j_c\propto  g_0$ since the magnitude of the order parameter stays roughly constant. As $g_0$ is increased, eventually the order parameter symmetry changes, making $j_c$ depart from the linear dependence, reaching a maximum until it monotonically decreases for large $g_0$.

\section{Discussion}

 The phase diagram of the twisted bilayer cuprate  provides us with insight into the nature of the TRSB state inferred from data in Ref.~\cite{FrankZhao}. We observed that the existence and nature of the  TRSB state is sensitive to the magnitude of the tunneling and doping of the lattice model.  We comment here on some possible numerical artifacts, and on how this model may be improved in the future.

 Continuum model calculations\cite{PhysRevB.105.064501,PixleyVolkov_review} also observe various types of TRSB states, which have been labeled in these works by their in-plane momentum dependence ($d+id$ or $d+is$). While these states are similar to the ones we also find, the notation used obscures somewhat  their dependence on layer index which we have discussed in detail.  To our knowledge, these works also did not explore the larger tunneling regime, where we find the fully gapped $A_1$ phase.  However, they have done an extensive exploration of the effects of temperature on the TRSB states.  Our spot checks of the $T$ dependence of a few cases appears to agree qualitatively with the earlier work in parameter regimes where comparisons are possible.  We performed our calculations starting with a random complex initial guess for the order parameter symmetry allowing for all irreps of the $D_4$ group as solutions whereas previous works have started with a $d+e^{i\phi}d^\prime$ ansatz \cite{Fidrysiak2023,Can2021}.
\par
The discrepancy in the experiments, Refs. \cite{FrankZhao} and \cite{Xikunxue}, may be attributed to how the Josephson junctions  were made. In Ref.~\cite{FrankZhao}, bulk BSSCO crystals were cleaved, twisted, and pressed together to form the junctions, whereas in Ref.~\cite{Xikunxue} the BSSCO flakes were placed on a substrate.  As mentioned in Ref.~\cite{Xikunxue}, there are variations in the tunnel barrier thickness due to local distortions.  Zhu et al. \cite{Xikunxue} speculated  that these distortions could allow for an $s$-wave component, but did not comment on a mechanism. We have pointed out here that enhanced tunneling will push the system locally towards an $s$ wave component or even pure $s$ ground state.
Zhu et al. further note the possibility of a spontaneous breaking of the $C_4$ lattice symmetry by strain, potentially giving rise to an $s$-wave component. This effect may be more pronounced in  thin flakes because fewer layers would offer less rigidity to shifts in positions of atoms in the unit cell. Finally, thin flakes are more susceptible to impurity effects which can lead to variations in the tunneling matrix elements\cite{Xikunxue}. 
It is evident from the above discussion that the tunneling form and strength play an important role in determining the order parameter symmetry.  We believe that the most likely source of discrepancy between Refs. \cite{FrankZhao} and \cite{Xikunxue} is the induced change in electronic structure near points of low tunneling barriers, driving the system locally in to a state with an $s$ wave component at 45$\degree$.
\par
Since tunneling plays such an essential role in the phase diagram,  it is important to perform a more precise treatment for the form of the tunneling matrix $g_{ij}$  in a realistic model of a twisted BSSCO bilayer. This may be accomplished by numerically calculating overlaps of surface Wannier functions\cite{Kreisel2015} at arbitrary twist angles. We will attempt a calculation of this type in future work.
\par

\section{Conclusions} 
 To map out the phase diagram of twisted bilayer cuprates, in this work we studied a mean field model of two square lattices twisted to  a relative commensurate angle, including in-layer nearest and next nearest hopping, nearest neighbor attraction and an interlayer tunneling term. We evaluated the superconducting  gap function, a complicated function of moir\'e coordinates, self consistently and determined the momentum and real space dependence.
\par
The moir\'e system formed by the commensurate twist of the two square lattices has a lower symmetry group $D_4$ as compared to the $D_{4h}$ group for the isolated square lattices. Using the basis functions for each irrep, we determined the symmetry of the converged order parameter across our phase diagram. We also use properties of the real space order parameter to verify TRSB behavior of the order parameter.
A thorough scan of the phase space as a function of twist angle, filling and doping demonstrates the presence of patches both TRSB and TRS states across the phase diagram. A study of the normal state DOS for $\theta=0\degree$ shows a strong correlation of the TRSB state with proximity of the Van Hove peaks to the chemical potential.  Thus doping and tunneling strength influenced the ground state via the Van Hove position.  However, we discussed also the role of the Fermi surface, which becomes more complex as the tunneling strength increases.  We further exhibited the importance of the high spectral weight portions of the Fermi surface to the antinodal regions for the various competing states.
\par
 Finally, we calculated the Josephson critical current for various states.  For small tunneling strength and filling near the Van Hove singularity, we reproduced the strong suppression of the critical current predicted by Can et al.\cite{Can2021} and measured by Zhao et al.~\cite{FrankZhao} near $45\degree$.  
However, the critical current at general nonzero twist  remains significant  for larger tunneling across a wide filling range in our calculations, because different TRSB states are dominant.  At even larger tunneling, trivial $A_1$ states are induced that exhibit generic critical currents. We discussed the possibility that these differences could account for the roughly angle-independent critical current observed in the monolayer twist experiments of Zhu et al.~\cite{Xikunxue}.  Clearly it is essential to understand the microscopic nature tunneling matrix elements between the BSCCO layers to confirm the validity of this hypothesis.  
\begin{acknowledgments}
The authors are grateful to M. Franz and O. Can for discussions regarding their calculations, and to P. Volkov and J. Pixley for valuable discussions.   We further thank Y. Wang for advice regarding the identification of TRSB states. S.P. and P.J.H. were  partially supported by
NSF-DMR-2231821.
A.K. acknowledges support by the Danish National Committee for Research Infrastructure (NUFI) through the ESS-Lighthouse Q-MAT.
\end{acknowledgments}

\appendix

\section{Algorithm}
\label{algorithm}
The Hamiltonian for the twisted bilayer cuprate model is defined in Eq.~(\ref{eq:Hamiltonian}). 
We need to redefine our Hamiltonian in the moir\'e setup. Each site in our lattice is relabeled using two indices $i\rightarrow(\alpha,\mathbf{u})$ where $\alpha$ is the site index inside the moir\'e unit cell and $\mathbf{u}=(u_x,u_y)$ is the moir\'e unit cell coordinate. We also get rid of the layer index by labeling all the  $2(m^2+n^2)$ sites in our moir\'e unit cell  for a commensurate angle $(m,n)$ as $0,1,...,2(m^2+n^2)$ and since each monolayer has $m^2+n^2$ sites if $\alpha\leq(m^2+n^2)$ it belongs to first monolayer and if $(m^2+n^2)< \alpha\leq 2(m^2+n^2)$ then it belongs to the second monolayer. 
\par
\begin{figure}[tb]
   \centering
    \includegraphics[width=0.8\linewidth]{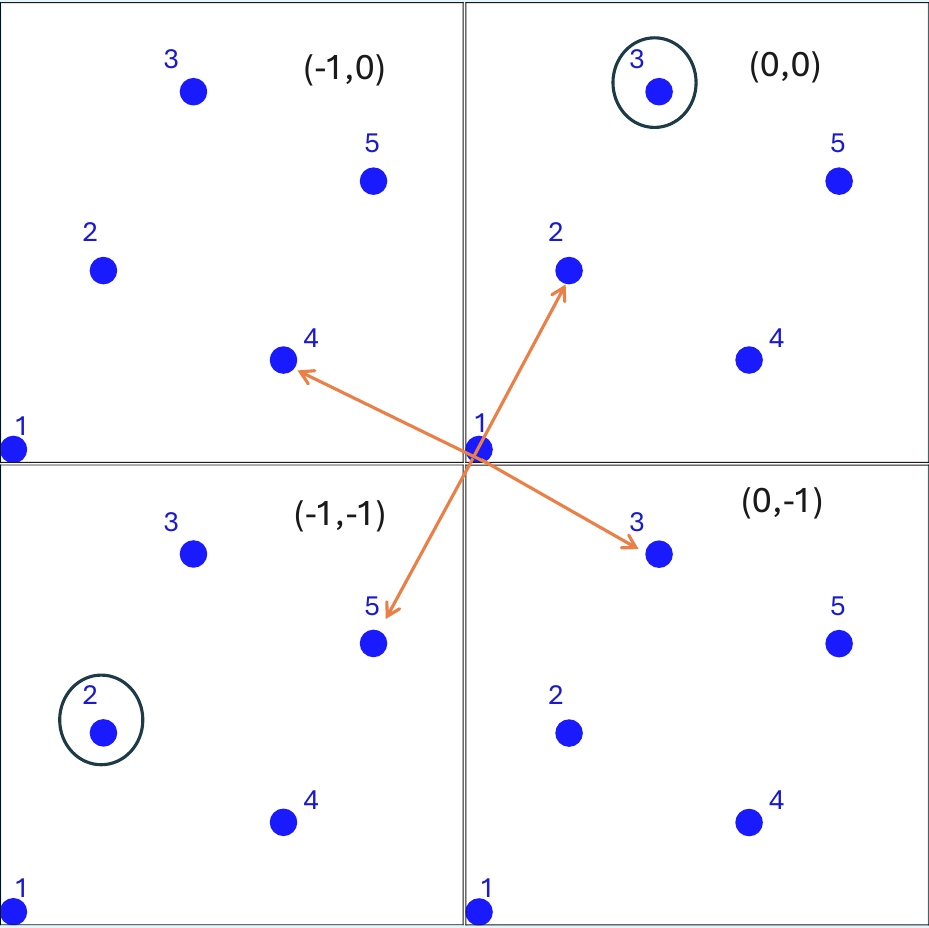}
    \caption{Nearest neighbor hopping terms for site 1 for $\theta=36.87\degree$. We show the sites in one monolayer for the moir\'e unit cells to which site 1 in unit cell (0,0) can hop using the orange arrows. The moir\'e unit cell coordinates are mentioned in the top right corner of each unit cell. Taking note of the site indices as well as the moir\'e unit cell coordinates provides us with all the information required to solve the Hamiltonian Eq.~(\ref{eq:HamiltonianMoireMF}). }
    \label{NNlist}
\end{figure}

We can understand the above notation using an example for $\theta=53.13\degree$. For this case $(m,n)=(1,2)$ and the moir\'e unit cell is shown in Fig.~\ref{supercell}. The index runs from $1,...,10$ where $1...5$ marked with blue belong to  the bottom monolayer and $6...10$ marked with red belong to the top monolayer. In Fig.~\ref{NNlist} we have neighboring moir\'e unit cell for the bottom monolayer and the upspin creation operator corresponding to the sites marked by black circles are  $c_{3\uparrow}^{(0,0)\dagger}$ and  $c_{2\uparrow}^{(-1,-1)\dagger}$.
\par
We rewrite the Hamiltonian in Eq.~(\ref{eq:Hamiltonian}) in the moir\'e setup
\begin{equation}
\begin{aligned}
H = & -\sum_{(\alpha,\mathbf{u}),(\beta,\mathbf{v}),\sigma } t_{\alpha\beta}^{\bf u\bf v} c_{\alpha\sigma }^{\bf u\dagger} c_{\beta\sigma} ^{\bf v} 
- \mu \sum_{(\alpha,\bf u),\sigma} n_{\alpha\sigma }^{\bf u}\\
& - \sum_{(\alpha,\bf u),(\beta,\bf v)} V_{\alpha\beta}^{\bf u \bf v} n_{\alpha}^{\bf u} n_{\beta} ^{\bf v} 
- \sum_{(\alpha,\bf u),(\beta,\bf v)\sigma} g_{\alpha\beta}^{\bf u \bf v} c_{\alpha\sigma}^{\bf u\dagger} c_{\beta\sigma}^{\bf v}\ 
\end{aligned}
\label{eq:HamiltonianMoire}
\end{equation}
\par
To evaluate the SC gap equation we first need to perform a mean field decomposition of Eq.~(\ref{eq:HamiltonianMoire}) resulting in
\begin{equation}
\begin{aligned}
H = & -\sum_{(\alpha,\bf u),(\beta,\bf v),\sigma } t_{\alpha\beta}^{\bf u \bf v} c_{\alpha\sigma }^{\bf u\dagger} c_{\beta\sigma}^{\bf v} 
- \mu \sum_{(\alpha,\bf u),\sigma} n_{\alpha\sigma }^{\bf u}\\
& + \sum_{(\alpha,\bf u),(\beta,\bf v)} \Delta_{\alpha\beta}^{\bf u \bf v} c_{\alpha\uparrow}^{\bf u\dagger} c_{\beta\downarrow} ^{\bf v\dagger} 
- \sum_{(\alpha,\bf u),(\beta,\bf v)\sigma} g_{\alpha\beta}^{\bf u \bf v} c_{\alpha\sigma}^{\bf u\dagger} c_{\beta\sigma}^{\bf v}\ ,
\end{aligned}
\label{eq:HamiltonianMoireMF}
\end{equation}
where
\begin{equation}
    \Delta_{\alpha\beta}^{\bf u\bf v} =V_{\alpha\beta}^{\bf u \bf v}\langle c_{\alpha\uparrow}^{\bf u}c_{\beta\downarrow }^{\bf v}\rangle .
    \label{eq:SC}
\end{equation}

We solve our Hamiltonian in momentum space and hence we have to perform the Fourier transform of the above Hamiltonian where the annihilation operators in the moir\'e setup transform as 
\begin{equation}
\begin{aligned}
c_{\alpha\uparrow}^{\bf u}=\sum_k c_{\alpha\mathbf{k} \uparrow}e^{i\mathbf{k}.\bf u} .
\label{eq:FT}
\end{aligned}
\end{equation}

We make a list of all possible hopping, tunneling and interaction processes which are  used to define the Hamiltonian. In the moir\'e setup while considering any of the inter site hopping, tunneling or interaction terms of the Hamiltonian we need to store the information about the site indices as well as the moir\'e unit cell coordinates as evident from Eq.~(\ref{eq:HamiltonianMoireMF}). We explain this using an example of $36.87\degree$ twist angle.  Since hoppings are defined in a monolayer we show the moir\'e unit cell with sites for one layer along with some of its neighboring cells which will be relevant in the further discussion in Fig.~\ref{NNlist}. The nearest neighbor hoppings from site 1 are shown in the figure. We can observe that the nearest neighbors for site 1  are site 2 in moir\'e unit cell (0,0), site 3 in moir\'e unit cell (0,-1), site 5 in moir\'e unit cell (-1,-1) and site 4 in moir\'e unit cell (-1,0).
\par

An extensive list of all possible hoppings will contain this information for all the sites in our moir\'e unit cell along with the hopping magnitude. We can generate similar list for next nearest neighbor hoppings as well as nearest neighbor interactions. The list of all possible tunneling terms can also be made similarly but the tunneling magnitude Eq.~(\ref{eq:tunnel}) changes for each $i,j$.

For any commensurate angle $(m,n)$ we can rewrite the BdG Hamiltonian in Eq.~(\ref{eq:HamiltonianMoireMF}) using the Nambu vectors defined as $[c_{1 k\uparrow},...,c_{2(m^2+n^2)k\uparrow},c_{1 -k\downarrow}^\dagger,...,c_{2(m^2+n^2) -k\downarrow}^\dagger]^{T}$. The moir\'e unit cell coordinate indices are not present here because they contribute in the Fourier transform as seen in Eq.~(\ref{eq:FT})
. The Hamiltonian in the matrix form is
\begin{equation}
 H_{BdG}(\mathbf{k})=
\begin{bmatrix}
H_0(\mathbf{k})& \hat{\Delta}(\mathbf{k})\\
\hat{\Delta}^\dagger(\mathbf{k}) & -H_0^T(-\mathbf{k})
\end{bmatrix}
,
\label{Eq.Hbdg}
\end{equation}
where $H_0(\mathbf{k})$ and $\hat{\Delta}(\mathbf{k})$ are $2(m^2+n^2)\times2(m^2+n^2)$ matrices.  $H_0(\mathbf{k})$ is the non-interacting part of the Hamiltonian consisting of hopping and tunneling terms whereas $\hat{\Delta}(\mathbf{k})$ contains the SC term of the Hamiltonian.
\par
The structure of the non-interacting part of the Hamiltonian is as follows
\begin{equation}
 H_0(\mathbf{k})=
\begin{bmatrix}
H_{\text{hop},1}(\mathbf{k}) & H_{\text{tun},12}(\mathbf{k})\\
H_{\text{tun},21}(\mathbf{k}) & H_{\text{hop},2}(\mathbf{k})
\end{bmatrix}  ,
\label{eq:non interacting Hamiltonian}
\end{equation}
where  $H_{\text{hop},1}(\mathbf{k})$ and $H_{\text{hop},2}(\mathbf{k})$ are $(m^2+n^2)\times(m^2+n^2)$ matrices for the hopping terms in first and second monolayer respectively and  $H_{\text{tun},12}$ and $H_{\text{tun},21}$ are also $(m^2+n^2)\times(m^2+n^2)$ matrices for the tunneling terms from layer $1\rightarrow 2$ and $2 \rightarrow1$ respectively. 
\par
 We explicitly show the Fourier transform of the  nearest neighbor hopping term in the moir\'e setup to obtain the corresponding matrix element using our Nambu vectors. We have also used the fact that all our nearest neighbor hopping have the same magnitude $t$.
 
\begin{align}
- &\sum_{\langle (\alpha,\bf u),(\beta,\bf v)\rangle \sigma }t_{\alpha\beta}^{\bf u \bf v} c_{\alpha \sigma }^{\bf u\dagger} c_{\beta \sigma }^{\bf v}&\notag \\
=& -t \sum_{\langle (\alpha,\bf u),(\beta,\bf v)\rangle \sigma  \mathbf{k} \mathbf{k}^{\prime}} c_{\alpha\mathbf{k} \sigma }^{\dagger} c_{\beta\mathbf{k}^{\prime} \sigma }e^{-i\mathbf{k}\cdot \mathbf u+i\mathbf{k}^{\prime}\cdot \bf v}&\notag \\
=&-t \sum_{\langle (\alpha,\bf u),(\beta,\bf v)\rangle \sigma  \mathbf{k} \mathbf{k}^{\prime}} c_{\alpha\mathbf{k} \sigma }^{\dagger} c_{\beta\mathbf{k}^{\prime} \sigma }e^{-i\mathbf{k}\cdot(\bf u-\bf v)}\delta_{\mathbf{k}\mathbf{k}^\prime}&\notag \\
=&-t \sum_{\langle (\alpha,\bf u),(\beta,\bf v)\rangle \sigma  \mathbf{k} } c_{\alpha\mathbf{k} \sigma }^{\dagger} c_{\beta\mathbf{k} \sigma }e^{-i\mathbf{k}\cdot(\bf u-\bf v)}.&
\label{eq:FTNN}
\end{align}
The list of hopping and tunneling terms which we had discussed earlier in this section can then be used to find the correct Fourier factors between any such hopping term in $H_{\mathrm{hop},1}(\mathbf{k})$ and $H_{\mathrm{hop},2}(\mathbf{k})$. For example- for $\theta=36.87\degree$ the nearest neighbor contribution for $H_{\mathrm{hop},1}^{12}(\mathbf{k})=-t$ since they both belong to the same moir\'e unit cell but $H_{\mathrm{hop},1}^{13}(\mathbf{k})=-te^{-ik_y}$ since the hopping connects the two moir\'e unit cells with coordinates $(0,0)$ and $(0,-1)$. We can find the contributions from the next nearest neighbor hopping and tunneling processes using the exact same method.
\par
As mentioned earlier the SC part of the Hamiltonian $\hat{\Delta}(\mathbf{k})$ is also a $2(m^2+n^2)\times2(m^2+n^2)$ matrix whose structure is greatly simplified because we only have interaction within each monolayer leading to offdiagonal elements being 0. The matrix is written as-
\begin{equation}
 \hat{\Delta}(\mathbf{k})=
\begin{bmatrix}
    \hat{\Delta}_1(\mathbf{k}) & 0\\
    0 & \hat{\Delta}_2(\mathbf{k})
\end{bmatrix}   ,
\end{equation}
where $\hat{\Delta}_1(\mathbf{k})$ and $\hat{\Delta}_2(\mathbf{k})$ are $(m^2+n^2)\times(m^2+n^2)$ matrices, evaluated for monolayer 1 and 2 respectively using the list of interactions we have discussed above  and Eq.~(\ref{eq:SC}) after performing Fourier transform as shown in Eq.~(\ref{eq:FTNN}). Since we only have nearest neighbor interactions we can rewrite
\begin{equation}
    \Delta_{\alpha\beta}^{\bf u \bf v} =V\langle c_{\alpha\uparrow}^{\bf u}c_{\beta\downarrow }^{\bf v}\rangle,
    \label{eq:SCnew}
\end{equation}
where $(\alpha,\bf u)$ and $(\beta,\bf v)$ are nearest neighbors. we define $\bf r=\bf u-\bf v$ and the Fourier transform is
\begin{equation}
\begin{aligned}
\Delta_{\alpha\beta}(\mathbf{k})&=\sum_{\bf r}V\langle c_{\alpha\uparrow}^{\bf u}c_{\beta\downarrow }^{\bf v}\rangle e^{-i\mathbf{k}\cdot \bf r}\\
&=\sum_{\bf r\mathbf{k}^{\prime}}V\langle c_{\alpha \mathbf{k}^\prime\uparrow}c_{\beta -\mathbf{k}^{\prime}\downarrow }\rangle e^{-i(\mathbf{k}-\mathbf{k}^\prime)\cdot \bf r}\\
&=\sum_{\bf k^\prime}V^{\mathbf k \mathbf k\prime}\langle c_{\alpha \mathbf k^\prime\up}c_{\beta-\mathbf k^\prime\down}\rangle,\\
\end{aligned}
\label{eq:FTSCOP}
\end{equation}
where
\begin{equation}
\begin{aligned}
V^{\mathbf k \mathbf k\prime}=\sum_{\bf r}Ve^{-i(\mathbf{k} -\mathbf{k}^\prime)\cdot \bf r}.
\end{aligned}
\end{equation}
\par
For a given twist angle and other Hamiltonian parameters, the calculation starts with defining a random complex initial guess for $\hat{\Delta}(\mathbf{k})$ for a defined momentum grid and evaluating the non-interacting part of the $H_{BdG}$ Hamiltonian (Eq.~(\ref{Eq.Hbdg})) using the model parameters. We diagonalize $H_{BdG}$ to obtain the quasiparticle eigenvalues and eigenvectors. The eigenvector matrix is inverted to obtain the creation and annihilation operators in our Nambu vectors in terms of the quasiparticle states. Using the creation and annihilation operators we  calculate the filling $\nu=\sum_{\alpha\mathbf{k}\sigma}\langle c_{\alpha\mathbf{k}\sigma}^{\dagger}c_{\alpha\mathbf{k}\sigma} \rangle$ to evaluate the chemical potential self-consistently and also  $\hat{\Delta}(\mathbf{k})$ which is then used as the input for the next iteration. The procedure is repeated till the order parameter converges to the desired precision.

\section{The analytical form of the order parameter for $\theta=90\degree$}
\label{90derive}
The $90 \degree$ twist angle case is physically identical to the $0\degree$ twist case because of the $C_4$ symmetry of our twisted bilayer model. The difference between the two cases are in the moir\'e unit cell since $0\degree$  has two sites, one per monolayer in the unit cell whereas $90\degree$ has 4 sites, 2 per monolayer. This difference leads to different momentum dependence of the order parameter in the two cases and also has a non-trivial imaginary part for $90\degree$.
\par 
We can derive the analytic expression of the order parameter by starting with
\begin{equation}
    H_{\Delta}=\sum_{(\alpha,\bf u),(\beta,\bf v)}\Delta_{\alpha\beta}^{\bf u \bf v}c^{\bf u\dagger}_{\alpha\uparrow }c^{\bf v\dagger}_{\beta\downarrow } + h.c.,
    \label{Ham90}
\end{equation}
where $H_\Delta$ is the SC part of the Hamiltonian, $\Delta_{\alpha\beta}^{\bf u \bf v}$ is the order parameter between sites $(\alpha,\bf u),(\beta,\bf v)$.
\par
\begin{figure}[tb]
   \centering
    \includegraphics[width=0.8\linewidth]{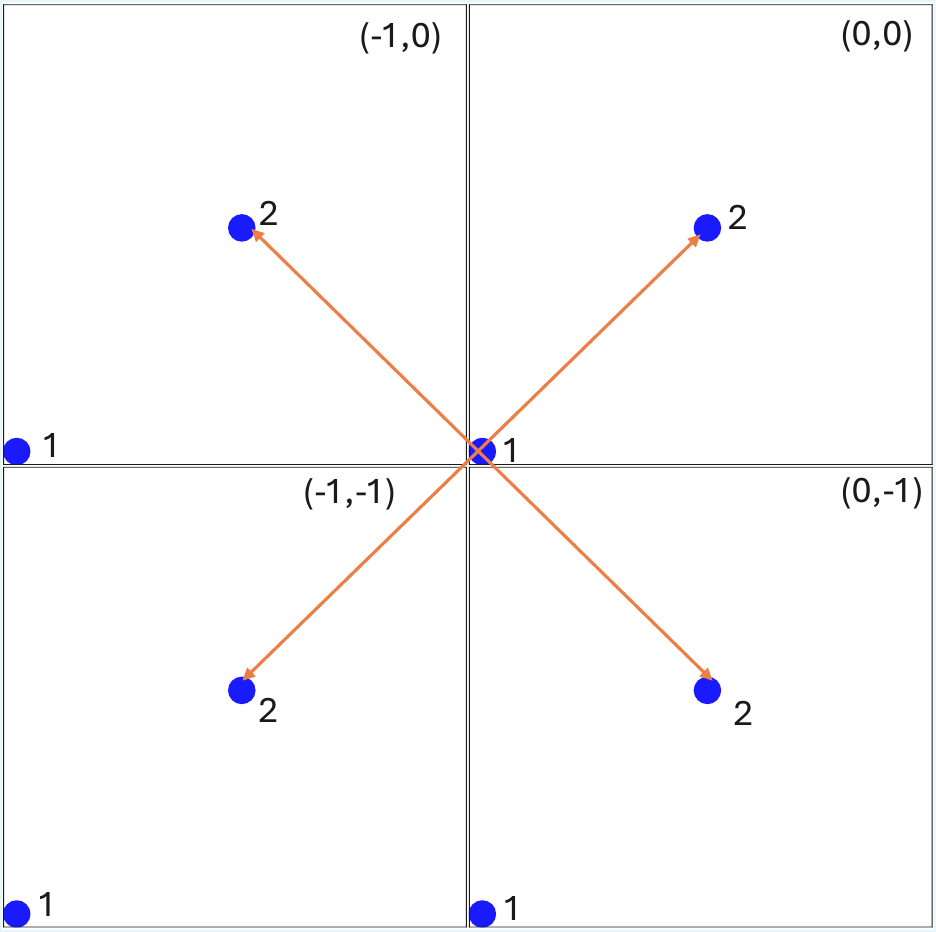}
    \caption{Nearest neighbor hopping terms for site 1 for $\theta=90\degree$. We show the sites in one monolayer for the moir\'e unit cells to which site 1 in unit cell (0,0) can hop using the orange arrows. The moir\'e unit cell coordinates are mentioned in the top right corner of each unit cell. }
    \label{90NNlist}
\end{figure}

We need to perform the Fourier transform of Eq.~(\ref{Ham90}) using the list of nearest neighbor interactions. We perform the calculation for $\Delta_{12}(\mathbf{k})$ term in the order parameter matrix as an example. Fig.~\ref{90NNlist} displays the nearest neighbors for site 1. The list includes site 2 belonging to moir\'e unit cells with coordinates \mbox{$(0,0)$, $(0,-1)$, $(-1,0)$, $(-1,-1)$} in units of moir\'e unit cell lattice constant $b$.

 \begin{figure}[tb]
 \includegraphics[width=\linewidth]{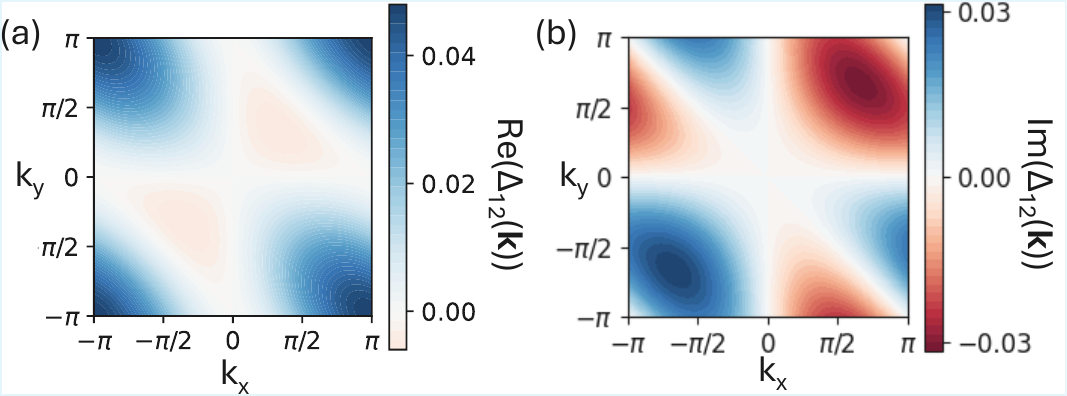}
\caption{Analytic calculation results for $\theta=90\degree$. (a) Real part of $d$ wave order parameter. (b) Imaginary part of $d$ wave order parameter. The from exactly matches the $\Delta_{12}$ form obtained in the numerical results in Fig.~\ref{numericalreal11} }
\label{90 angle d wave analytical}
 \end{figure}

\begin{figure*}[tb]
 \includegraphics[width=\linewidth]{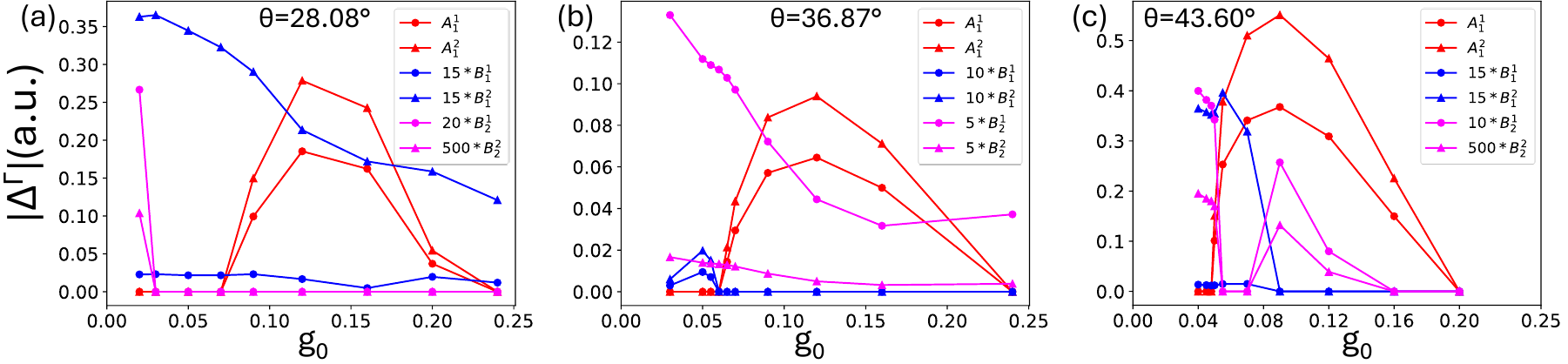}
         \caption{$|\Delta^{u}|$ vs $g_0$ for various $\theta$. (a) $\theta=28.08\degree$ and $\nu=0.49$. (b) $\theta=36.87\degree$ and $\nu=0.45$. (c) $\theta=43.6\degree$ and $\nu=0.4$. The notation of the basis function in the plot legends has been provided in Table.~\ref{CharacterTableD4}. For different $\theta$ we have different normalization factors for the irreps so that they lie on the same scale (see legend). }
         \label{Opvsgo}
 \end{figure*}
Expanding Eq.~(\ref{Ham90}) for the above list of hoppings and performing the Fourier transform gives us
\begin{align}
H_{\Delta_{12}}&=\sum_{\bf {k},\bf {k'}}\Delta_{12}^{(0,0), (0,0)}c^{\dagger}_{1\bf k\uparrow } c^{\dagger }_{2\bf k'\downarrow } e^{-i\mathbf{k}\cdot(0,0) }e^{-i\mathbf{k'}\cdot (0,0)}\nonumber\\
&+\Delta_{12}^{(0,0), (0,-1)}c^{\dagger}_{1\bf k\uparrow } c^{\dagger }_{2\bf k'\downarrow } e^{-i\mathbf{k}\cdot(0,0) }e^{-i\mathbf{k'}\cdot (0,-1)}\\
&+\Delta_{12}^{(0,0), (-1,-1)}c^{\dagger}_{1\bf k\uparrow } c^{\dagger }_{2\bf k'\downarrow } e^{-i\mathbf{k}\cdot(0,0) }e^{-i\mathbf{k'}\cdot (-1,-1)}\nonumber\\
&+\Delta_{12}^{(0,0), (-1,0)}c^{\dagger}_{1\bf k\uparrow } c^{\dagger }_{2\bf k'\downarrow } e^{-i\mathbf{k}\cdot(0,0) }e^{-i\mathbf{k'}\cdot (-1,0)}+h.c.
\nonumber
\end{align}    

For a $d$ wave order parameter bonds at an angle $90\degree$ from each other have opposite signs and equal magnitude. Using this property our equation is
\begin{equation}
 \begin{aligned}
H_{\Delta_{12}}&=\sum_{\bf {k},\bf {k'}}\Delta_{12}c^{\dagger}_{1\bf k\uparrow } c^{\dagger }_{2\bf k'\downarrow } e^{-i\mathbf{k}\cdot(0,0) }e^{-i\mathbf{k'}\cdot (0,0)}\\
&-\Delta_{12}c^{\dagger}_{1\bf k\uparrow } c^{\dagger }_{2\bf k'\downarrow } e^{-i\mathbf{k}\cdot(0,0) }e^{-i\mathbf{k'}\cdot (0,-1)}\\
&+\Delta_{12}c^{\dagger}_{1\bf k\uparrow } c^{\dagger }_{2\bf k'\downarrow } e^{-i\mathbf{k}\cdot(0,0) }e^{-i\mathbf{k'}\cdot (-1,-1)}\\
&-\Delta_{12}c^{\dagger}_{1\bf k\uparrow } c^{\dagger }_{2\bf k'\downarrow } e^{-i\mathbf{k}\cdot(0,0) }e^{-i\mathbf{k'}\cdot (-1,0)}+h.c.\\
 \end{aligned}   
\end{equation}
In the Cooper channel $\mathbf{k}^\prime=-\mathbf{k}$ and hence we obtain.
\begin{equation}
 \begin{aligned}
H_{\Delta_{12}}&=\sum_{\bf {k}}\Delta_{12}c^{\dagger}_{1\bf k\uparrow } c^{\dagger }_{2\bf -k\downarrow } \\
&-\Delta_{12}c^{\dagger}_{1\bf k\uparrow } c^{\dagger }_{2\bf -k\downarrow } e^{-ik_y}\\
&+\Delta_{12}c^{\dagger}_{1\bf k\uparrow } c^{\dagger }_{2\bf -k\downarrow } e^{-i(k_x+k_y)}\\
&-\Delta_{12}c^{\dagger}_{1\bf k\uparrow } c^{\dagger }_{2\bf -k\downarrow } e^{-ik_x}+h.c,\\
 \end{aligned}   
\end{equation}
and thus the Hamiltonian reduces to
\begin{align}
    H_{\Delta_{12}}(\bf k)&=\Delta_{12}c^{\dagger }_{1\bf k\uparrow } c^{\dagger }_{-2\bf k\downarrow } (1-e^{ik_y}+e^{i(k_x+k_y)}-e^{ik_x}) \nonumber\\
     +h.c
\end{align}
giving us an order parameter of the form 
\begin{align}
    \Delta_{12}(\bf k)&= \mathop{\Delta_{12}}[(1+\cos(k_x+k_y)\notag
    -\cos(k_x)-\cos(k_y)\notag)\\
	&+i(\sin(k_x+k_y)-\sin(k_y)-\sin(k_x))].
\label{11 d wave appendix}
\end{align}
which we plot in Fig.~\ref{90 angle d wave analytical}.

We can similarly find the order parameter form for the  matrix elements of other twist angles $(m,n)$ but it gets analytically challenging because  the number of matrix elements in the order parameter matrix increase as we reach closer to $45\degree$ as $m$ and $n$ also increase 
\section{Order parameter evolution}
\label{Opevol}

The phase diagram in Fig.~\ref{Phase diagram} displays the various states and transitions between them. To clearly observe these features we plot the  non-normalized magnitudes of the order parameters$(|\Delta^{\Gamma}|)$ in those phases, in Fig.~\ref{Opvsgo}. We project the converged order parameter onto different irreps basis functions  and examine its evolution {\it vs} $g_0$. The order parameter in these plots for different $\theta$ have been scaled differently for visual clarity (see legend).  For $\theta=28.08\degree$ and $\nu=0.49$ Fig.~\ref{Opvsgo}(a) shows that the $B_1$ order parameter  is present throughout, whereas the $B_2$ order parameter is non-zero yet smaller then the $B_1$ state until it reaches 0 for $g_0=0.03$ indicating a $B_1+iB_2$ state for small $g_0$. Between $g_0\approx[0.03,0.07]$ a $B_1$ wave state is observed, and the $A_1$ order parameter with a larger magnitude coexists with the $B_1$ wave state for $g_0\approx[0.07,0.24]$, indicating a $B_1+iA_1$ state. For $\theta=36.87\degree$ and $\nu=0.45$  Fig.~\ref{Opvsgo}(b) shows a  $B_2$ order parameter persisting throughout, whereas the $B_1$ order parameter is small and vanishes at $g_0=0.06$ and the $A_1$ order parameter then persists from $g_0\approx[0.06,0.24]$ and is the largest component of the order parameter for almost the whole range. For $\theta=43.6\degree$ and $\nu=0.4$ shown in Fig.~\ref{Opvsgo}(c) we observe $B_1$, $B_2$ and $A_1$ having non-zero values for $g_0=0.05$ indicating the coexistence region. Unique features of this twist angle include the existence of the  gapped TRS $A_1$ state for large $g_0$ and the switching of the order parameter from $B_1$  to $B_2$ symmetry for $g_0=0.07$ but this is not a universal feature for all $\nu$. For $g_0>0.2$ superconductivity is not stabilized in this parameter regime. For some of the twist angles in Fig.~\ref{Opvsgo} we observe an order of magnitude difference between the scaling factors of two irreps but since the order parameter projections are not normalized for all possible basis functions we cannot make physical predictions.

\end{document}